\documentclass{aa}

\newcommand{\hi}{\ion{H}{i}}
\newcommand{\kms}{km\,s$^{-1}$}
\usepackage{graphicx}

\usepackage{txfonts}
\usepackage{natbib}
\usepackage{graphicx}

\begin{document}

\title{Galactic \hi\, supershells: Kinetic energies and possible origin \thanks{Tables 3 and 4 are only available in electronic form at the CDS via anonymous ftp to cdsarc.u-strasbg.fr (130.79.128.5) or via http://cdsweb.u-strasbg.fr/cgi-bin/qcat?J/A+A/}}

\titlerunning{Energetics involved in \hi\, supershells}

\author{L. A. Suad\inst{1}, C. F. Caiafa\inst{2,3,4}, S. Cichowolski\inst{1}, \and E. M. Arnal\inst{5}}

\institute{ Instituto de Astronom\'{i}a y F\'{i}sica del Espacio (CONICET-UBA), Ciudad Universitaria, C.A.B.A, Argentina.  
\and Instituto Argentino de Radioastronom\'{i}a (CCT-La Plata, CONICET; CICPBA), C.C. No. 5, 1894,Villa Elisa, Argentina.
  \and Facultad de Ingenier\'{i}a, Universidad de Buenos Aires (FIUBA), C.A.B.A, Argentina.
  \and Tensor Learning Unit - RIKEN Center for Advanced Intelligence Project, 1-4-1 Nihonbashi, Chuo-ku, 103-0027, Tokyo, Japan.
  \and Facultad de Ciencias Astron\'omicas y Geof\'{\i}sicas, Universidad Nacional de La Plata, La Plata, Argentina.
}

\date{\today}

\abstract{The Milky Way, when viewed in the  neutral hydrogen line emission, presents large structures called Galactic supershells (GSs). The  origin of these structures is still a  subject of debate. The most common scenario invoked  is the combined action of strong winds from massive stars and their subsequent explosion as supernova.}
{The aim of this work is to determine the origin of 490 GSs that belong to the  Catalog of \hi\, supershell candidates in the outer part of the Galaxy.} 
{ To know the physical processes that took place to create these expanding structures, it is necessary to determine their kinetic energies. To obtain all the GS masses, we  developed and used an automatic algorithm, which was tested on 95 GSs whose masses were also estimated by hand.} 
{  The estimated kinetic energies of the GSs vary from  $1 \times 10^{47}$ to $3.4 \times 10^{51}$ erg. Considering an efficiency of 20\% for the conversion of mechanical stellar wind energy into the kinetic energy of the GSs, the estimated values of the GS energies could be reached by stellar OB associations. For the GSs located at high Galactic latitudes, the possible mechanism for their creation could be attributed to  collision with high velocity clouds (HVC). We have also analysed the distribution of GSs in the Galaxy, showing that at low Galactic latitudes, $|b| < 2^\circ,$ most of the structures in the third Galactic quadrant seem to be projected onto the Perseus Arm. The detection of GSs at very high distances from the Galactic centre may be attributed to diffuse gas associated with the circumgalactic medium of M31 and to intra-group gas in the Local Group filament.}{}


   \maketitle
   
 \section{Introduction}\label{intro}

The interstellar medium (ISM) of galaxies like the Milky Way has quite a complex structure where  phases with different physical properties coexist. Immersed in this ISM are a plethora of features with a large variety of denominations like bubbles, shells, supershells, chimneys, worms, holes, and so forth, which have been observed to exist throughout the electromagnetic spectrum.
        Amongst the above features quite a few shells and their over-sized cousins, the supershells, are observable in the \hi\,-line emission at a wavelength of 21-cm. Though there are in the astronomical literature a vast number of examples of \hi\,-shells likely to be physically associated with massive stars and its descendants, and/or supernova remnants \citep[e.g.][]{arn92,arn99,cic01,cap02,mcc02,cic04,pin08,cic09,cap10, arn11,cic14,rey17},  this is not the case for the majority of the  catalogued Galactic supershells (GSs for short),  where we considered GSs to those structures whose linear size is larger than 200 pc.
  Due to the physical association between \hi\, shells and stars mentioned above, it is widely accepted that their genesis is likely to be deeply rooted in the interaction of the associated star(s) with its surrounding ISM, through the individual effects of their stellar winds or the combined action of stellar winds and the posterior supernova (SN) explosion of the massive stars.
Since in most of the GSs the stellar counterpart possibly associated with them has not been observed, a stellar origin for the GSs similar to the one put forward for the \hi\,-shells is far from being clear. 

Furthermore, since for a few GSs their kinetic energy (E = 0.5\, M$_{GS}\,v_{\rm exp}^2$) derived from the observations is very high, greater than 10$^{52}$ ergs \citep{hei79},  it is thus unlikely to be provided by the combined action of stellar winds and SN explosions, unless we are willing to accept the existence in the past of stellar aggregates, open clusters, and/or OB associations that have an upper mass end much more populated than similar objects  known to exist today in the Milky Way.
For these extreme cases other plausible alternatives, like events connected to a gamma-ray burst \citep{per00} or the in-fall of high velocity clouds (HVC) into the Galactic \hi\, disc \citep{ten81}, may be at work to create these large structures.
    
Knowledge of the kinetic energy of a sizable sample of GSs in our Galaxy may be a possible way to shed some light on the genesis of GSs. A "stellar option" for the genesis of GSs would be favourable if most of the GS’s mechanical energy is within the range that could be injected by the winds and SN explosions of the most massive members of the stellar aggregates, open clusters, and/or OB associations. Otherwise alternative options like those previously mentioned ought to be further explored.

Though at first glance to derive the kinetic energy of a GS may appear rather simple, only the GS’s total mass and its expansion velocity are needed, the derivation of the mass is far from being trivial. Different  researchers may apply different criteria for its derivation, resulting in mass estimates for the same object that may be quite different. Since the goal of this work is to determine in a systematic way the mechanical energy of a statistically significant sample of GSs, to this end a computer algorithm that automatically computes the GS masses was developed by one of us (CFC). This algorithm was applied to most of the   objects  catalogued as a GS by \citet{sua14}. Though this catalogue contains a total of 566 structures, the algorithm was applied only to those GSs showing \hi\,-emission surrounding  its central cavity in at least three quarters (or 270 degrees) of its angular extent. A total of 490 GSs fulfilled this criterion.

\section{Observations}

\hi\, data were retrieved form the Leiden-Argentine-Bonn (LAB) survey \citep{kal05}. This database has an angular resolution of 34', a velocity resolution of  1.3 \kms,  a channel separation of 1.03 \kms, and it covers the velocity range from $-400$ to $+450$ \kms. The entire database has been corrected for stray radiation \citep{kal05}.

 \section{Estimation of GS masses}

\begin{figure*}[h]
\centering
\includegraphics[width=14cm]{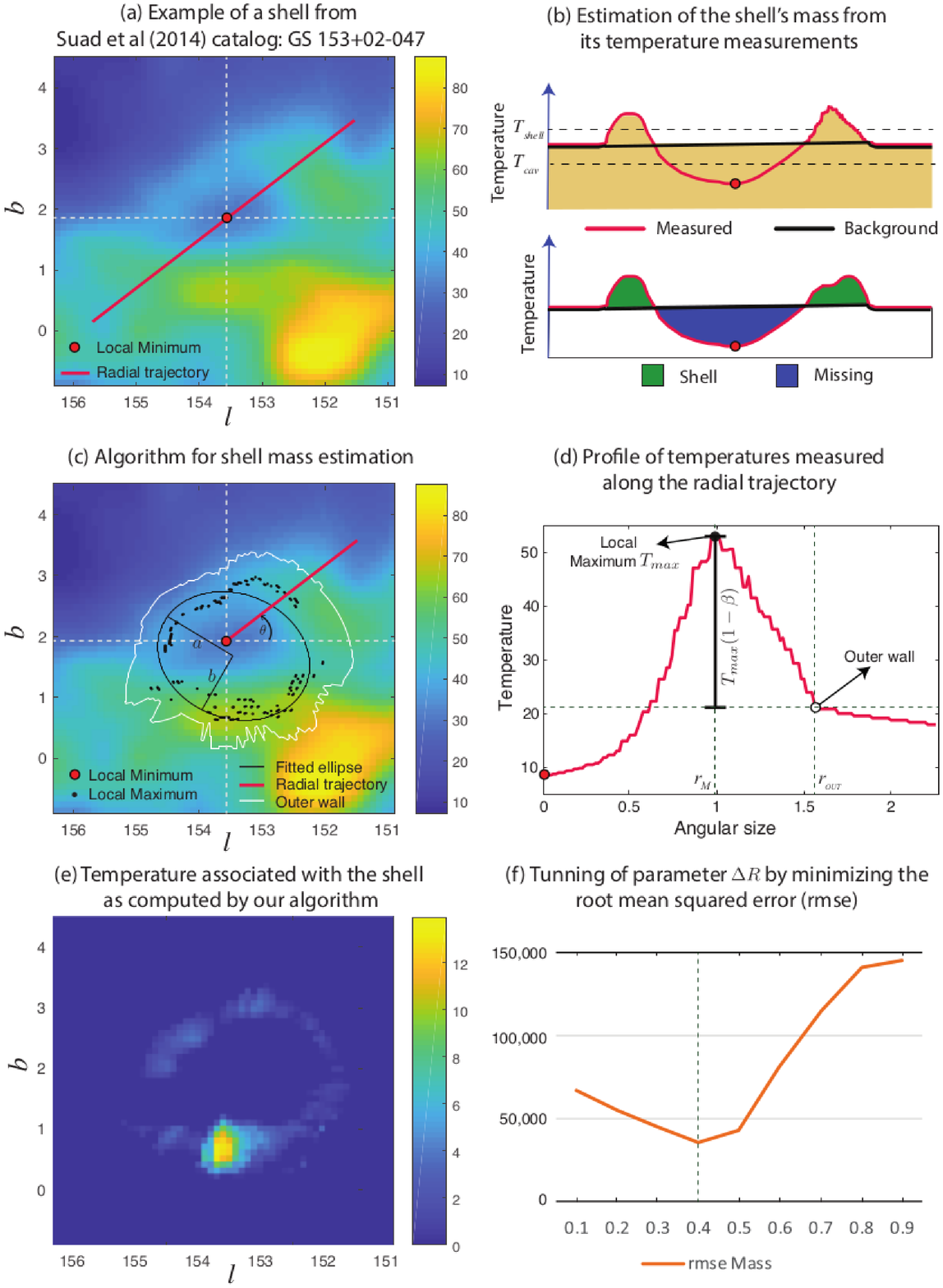}
\caption{Estimating the mass associated with a Galactic supershell: the example of GS\,153+02-047: 
(a) Averaged temperature image computed using the $70\%$ of channels where the GS is detected in order to avoid the caps; 
(b) Top: Temperature profile computed along the red line trajectory shown in (a). The background temperature profile (black) is defined by connecting with a line the external limits of the shell.  The averaged temperatures of the shell and the cavity are $T_{\rm shell}$ and $T_{\rm cav}$,  respectively. Bottom: Temperature of the shell (green) is computed as the excess of temperature with respect to the background temperature level. The temperature associated with the missing mass (blue) is computed as the required temperature to be added in the cavity in order to reach the background level; 
(c) Description of the algorithm for computing masses. Radial trajectories (red lines) are used for computing temperature profiles and detecting local maxima (black dots) and outer walls of the structure. Once local maxima are detected, the algorithm fits an ellipse to those points (black line); 
(d) Details of the temperature profile computed based on the radial trajectory displayed in panel (c);
(e) Image of the temperature of the shell as computed by subtracting the background from the measured temperature as explained in panel (b) and Eq. \ref{eq:Tshell};
(f) The global parameter $\Delta R$ is tuned such that the error of the algorithm (rmse) is minimized when it is compared against a subset of structures analysed "by hand". The optimal value of this parameter is $\Delta R=0.4$.
}
\label{fig:Fig_1}
\end{figure*}

As mentioned above, the GSs are  large voids  surrounded completely or partially by walls of \hi\, emission. An example of such a structure,  GS\,153+02-047, listed in the catalogue of \cite{sua14}, is shown in Fig. \ref{fig:Fig_1}a, along with a profile of temperatures measured along a scan that crosses the central minimum as shown in Fig. \ref{fig:Fig_1}b-top. Based on the hypothesis that before the structure was formed the \hi\, was uniformly distributed, we can assume that the excess mass, or temperature, in the shell should be equal to the mass missing in the void. In Fig. \ref{fig:Fig_1}b-bottom, the excess mass (shell mass, green) and missing mass (blue) are displayed for this particular example. 
Thus, we  can estimate both the excess mass of the \hi\, shell surrounding the cavity, denoted by  $M_{\mathrm{\hi\,}}^{\mathrm{shell}}$, and  the mass missing in the void, which we shall refer to as the missing mass, $M_{\mathrm{\hi\,}}^{\mathrm{miss}}$.
In an ideal case these two values should be exactly the same, however,
owing to inhomogeneities present in the ISM where the structures are located, the estimated values for both masses do not always match.

Calculation of the \hi\, masses  is straightforward, first
column densities are simply computed by integration of the
 difference in the mean brightness temperature ($\overline{T_{\rm b}}$) over the  velocity range  where the supershell is detected using

\begin{equation}
N_H = 1.823 \times 10^{18}\, \int{\overline{\Delta T_b}\, dv}\, \hspace{2.8cm} \, \, \, \rm [cm^{-2}]
\end{equation}

\noindent under the assumption that the \hi\, gas is optically thin.
Then, the neutral mass of an \hi\, structure of area $A$ is given by 

\begin{equation}
M_{\rm \hi\,} = N_H\, A
.\end{equation}

For a structure located at a distance  $d$  and  with an   angular size $\Omega$, $A = \Omega \, d^2$, and the mass
can be easily  estimated by

\begin{equation}
M_{\mbox{H\,\sc{i}}} = 1.3 \times 10^{-3}\, d_{\rm kpc}^2\, \Delta {\rm v}_{\rm km\,s^{-1}} \, \overline{\Delta T_{b}} \, \Omega\hspace{1.9cm} [M_{\odot}]
\label{eq:masa}
,\end{equation}

\noindent which gives an approximation to  an exact integral over velocity. We just take $\Delta {\rm v}_{\rm km\,s^{-1}}$ as the velocity interval over which the GS is visible. In this equation, $\Omega$ is expressed in arcmin$^2$ units.

\subsection{Estimation by hand}\label{hand}

In this section we describe the procedure used to estimate the \hi\, mass related to a structure by using Eq. \ref{eq:masa} and the visualization and manipulation data program: Astronomical Image Processing System (AIPS). 
The first step is to determine all the involved parameters for the objects under consideration:  distance,  velocity interval where the structure is observed, angular size of the feature, and the mean temperature brightness difference ($\overline{\Delta T_b}$). 

The distances of all the catalogued structures are given in \citet{sua14}.
The velocity interval where the supershell is better observed is determined by inspecting the \hi\, data-cube, and is used to create a velocity  averaged image.
This averaged  image is used to estimate the  angular areas of the cavity ($\Omega_{\rm cav}$) and the shell ($\Omega_{\rm shell}$), and their corresponding averaged temperatures, $\overline{T_{\rm cav}}$ and $\overline{T_{\rm shell}}$.
Since  $M_{\mathrm{\hi\,}}^{\mathrm{shell}}$ and $M_{\mathrm{\hi\,}}^{\mathrm{miss}}$ are the excess mass in the shell and the missing mass in the cavity with respect to what it was in the region before the structure was formed, we define $T_{\rm bg}$
as the background temperature, that is the temperature that  the  uniform ISM had before the mass was swept-up in a shell structure (see Fig. \ref{fig:Fig_1}b-top).
Then, using Eq. \ref{eq:masa}, we estimate the shell  masses by replacing $\overline{\Delta T_{b}}$ by ($\overline{T_{\rm shell}} - T_{\rm bg}$)  and $\Omega$ by $\Omega_{\rm shell}$; and the missing masses using  $\overline{\Delta T_{b}}$ = ($\overline{T_{\rm cav}} -T_{\rm bg}$)  and $\Omega =  \Omega_{\rm cav}$.

Though the background temperature can be estimated considering the temperature of the \hi\, emission located beyond the outer border of the GS, its determination is not straightforward. This stems from the fact that in most cases $T_{\rm bg}$ changes its value as we "move" around the outer part of the structure, due to variations  that are intrinsic to the overall \hi\, emission of the Milky Way.  
To determine it, we made several cuts in different directions of the structure and, from these profiles, we analysed what is the temperature of the gas emission not related to the structure.

All the estimated  values have  large uncertainties that must be taken into account in the value of the mass.
For instance, given the non-uniform background, the determination of both the shell outer limit and the background temperature is usually quite subjective and not easy.
To get an idea of the uncertainty involved in this measurement, the mass of each structure has been estimated by  two of the authors of this work (Suad and Cichowolski). Comparing the obtained values we have concluded that on average the difference is of the order of 50\%, both sides.

As another check of the accuracy of the estimated masses, we computed the background temperature local to each GS by forcing the missing and shell masses to be equal. The ratio between the temperature  obtained in this way, $T^*_{\rm bg}$, and the one obtained by inspecting each structure, $T_{\rm bg}$, is in the range between 1 and 1.2, indicating that the $T_{\rm bg}$ values used to compute the masses are in agreement with the hypothesis of $M_{\mathrm{\hi\,}}^{\mathrm{shell}}$ and $M_{\mathrm{\hi\,}}^{\mathrm{miss}}$  being equal.

\subsection{Algorithm for computing GS masses}
\label{algorithm}
In order to have a more precise estimation of the GS masses for a total of 490 structures included in the catalogue, we developed a computer-based algorithm and validated the results by comparing them against the values obtained "by hand" for a reduced subset of structures (see Sect. \ref{sec:validation}). 

The algorithm has three parameters that need to be tuned (see Sect. \ref{sec:tunning}). Two of them are structure-dependent parameters: the local maxima prune parameter $\alpha$, and the  outer wall extent parameter  $\beta $; and one global parameter (for all shells in the catalogue): the {\it maximum shell width} parameter $\Delta R$. The parameters are precisely defined in the algorithm description below. 

\subsubsection{Algorithm description}
For each of the structures in the catalogue, the algorithm performs the following steps:
\begin{itemize}
\item {\bf STEP 1 (Averaging):} The average map of temperatures, $T_b$, is obtained by averaging 70$\%$ of the central channels in which the shell is detected. This simple technique allows us to use a single averaged image to estimate the mass of the shell. This choice of the percentage of considered channels allowed us to successfully remove the "caps" of the structure making the minimum visible in the average.
Figure \ref{fig:Fig_1}a shows the averaged image obtained for the structure GS\,152+02$-$047.\\

\item {\bf STEP 2 (Local maxima finding):} 

{\bf I.} By setting the local minimum as the centre of the structure, the algorithm computes the temperature profiles $T(r,\theta)$ for 100 radial lines (trajectories) starting at the local minimum and moving outward, where $r$ is the angular distance from the centre and $\theta$ is the angle of the radial trajectory (red line in Fig. \ref{fig:Fig_1}c). For each of these profiles, the algorithm finds the peak temperature (local maximum) following the same criterion as used in the detection of the supershell candidates in \cite{sua14}. More specifically, in order to avoid non-realistic local maxima due, for example, to noise, we use the following criteria: first, we admit local maxima to exist only beyond a point of minimum slope. In other words, we compute the slope of the temperature profile and search for the local maximum only for distances beyond a point where a minimum slope threshold of $T_{slp} = 0.2K/px$ is exceeded. Second, a local maximum is defined in such a way that its brightness temperature exceeds by at least a threshold $\delta_T$ the brightness temperature of its surroundings. The value of $\delta_T = 0.4K$ was determined during the ``learning phase'' in our previous work \cite{sua14}. Third, we only accept a local maximum if its temperature exceeds the temperature at the minimum (centre of the shell) by a certain threshold. The value of this threshold depends on the position within the \hi\ data cube. In \cite{sua14}, we developed an interpolation technique to estimate an optimal threshold for every location in the data cube. Finally, the maximum temperature is denoted by $T_{max}=T(r_M,\theta)$, where $r_M$ is the distance from the minimum at which the maximum is attained. Figure \ref{fig:Fig_1}d shows a particular temperature profile corresponding to the angle $\theta$ as shown in Fig. \ref{fig:Fig_1}c.

{\bf II.} An ellipse is fitted to the local maxima points and the effective radii is defined as $R_{eff} = \sqrt{ab}$, where $a$ and $b$ are the semi-major and semi-minor axes of the fitted ellipse, respectively. Figure \ref{fig:Fig_1}c shows the detected local maxima (black points) for a particular structure in the catalogue together with the fitted elipse (black line).

{\bf III.} To avoid spurious local maxima, we filter out detected local maxima that are too far from the centre in relation to the rest of the points. Basically, we enforce the local maxima points 
to satisfy the  constraint
\begin{equation}
\label{eq:dist_max}
r_M < \alpha\, a,
\end{equation}
where $\alpha$ is the local maxima prune parameter to be determined and $a$ is the semi-major axis of the fitted ellipse. Once the points that do not meet the criterion in Eq. \ref{eq:dist_max} are removed, the algorithm repeats step II - III until convergence is reached.\\

\item {\bf STEP 3 (Outer wall determination):}
For each radial profile we need to determine the background temperature (solid black line in Fig. \ref{fig:Fig_1}b). To do so, we need to find the outer wall of the structure. The adopted criterion is to detect the point where the temperature drops by a factor $\beta \, T_{max}$. Formally we define $r^*$ such that $T(r^*)=\beta \, T_{max}$ , where $\beta$ is the {\it outer wall extent} parameter to be estimated (see below) and $T_{max}$ is the maximum temperature. Additionally, we impose the outer walls to be at a distance lower than a percentage $\Delta R$ of the effective radii computed in a previous step. Summarizing, we define the position $r_{out}$ of the outer wall such that
  
    \begin{equation}
       r_{out} = 
        \begin{cases}
            r^* & \text{if $r^* > R_{eff}\,\Delta R$} \\
            R_{eff}\, \Delta R & \text{otherwise}
        \end{cases}
    .\end{equation}
    
In Fig. \ref{fig:Fig_1}c the final shape of the outer wall is shown (white line). It is worth noting that, sometimes, for some ranges in the angle $\theta$  local maxima are not available and the algorithm finds the outer wall by interpolating the wall found for the nearest local maxima.\\

\item{\bf STEP 4 (Shell mass and missing mass estimation):}
Finally, for each of the radial lines  (each $\theta \in [0,\pi)$) where the temperature profiles were computed, we determine the background temperature  $T_{bg}(\theta,r)$, with $r\in (-r_{out}(-\theta),+r_{out}(\theta))$ by connecting the outer wall points in both ends of the profile as shown in Fig. \ref{fig:Fig_1}b (black line). Then, we obtain the  temperature of the shell  $T_{shell}(\theta,r)$ and the  "missing" temperature $T_{miss}(\theta,r)$ in every location $(\theta,r)$ of the shell as follows:
\begin{eqnarray}
\label{eq:Tshell}
T_{shell}(\theta,r) = h\,(T - T_{bg}) \\
T_{miss}(\theta,r) = h\,(T_{bg} - T),
\end{eqnarray}
where  the location $(\theta,r)$ in the right hand of the equations is avoided to simplify the notation, and $h(x)$ is the Heaviside step function, that is, $h(x) = x$ if $x \ge 0$ and $h(x) = 0$ otherwise.  It is noted that by combining the $T_{bg}(\theta,r)$ for all $\theta$ and $r$ in the shell, we finally obtained a surface of background temperatures, which allowed us to compute $T_{shell}(\theta,r)$ and $T_{miss}(\theta,r)$ in the shell. In Fig. \ref{fig:Fig_1}e, the resulting image for the temperature associated with the shell, $T_{shell}$, is displayed. Finally, the total temperature of the shell and the missing temperature are computed by integrating these temperature images in the plane $l-b$ and the associated shell and missing masses are computed through Eq. \ref{eq:masa}.

\end{itemize}

\subsubsection{Parameter tuning}
\label{sec:tunning}
To tune the parameters $\alpha$, $\beta,$ and $\Delta R$ we used a grid-search approach by running the algorithm for a wide range of parameter values ($\alpha \in [0.2,1.0]$, $\beta \in [0,0.75],$ and  $\Delta R \in [0.1, 0.9]$), computing the corresponding missing and shell masses for each case and choosing the optimal values according to the following criteria:
\begin{itemize}
\item {\bf Optimal choice of structure-dependent parameters $\alpha$ and $\beta$ given $\Delta R$}: We choose the value of $\alpha$ to be more in a way that the area covered by the wall is as close as possible to the area of the fitted ellipse, which was already estimated for each shell in the catalogue in \cite{sua14}. On the other side, the parameter $\beta$ is tuned by minimizing
the absolute difference between the $T_{shell}$ and $T_{miss}$ for a given $\Delta R$. We observed that the optimal value for these two parameters is different for each shell so we tune this parameter individually.

\item{\bf Optimal choice of global parameter $\Delta R$ (maximum shell width)}: Once parameters $\alpha$ and $\beta$ are tuned for each shell and each value of $\Delta R$ in a range, we choose the optimal $\Delta R$ as the one that minimizes the error in the estimation of the masses compared to the values obtained by hand, for a subset of 61 structures (Group $A$, see Sect. \ref{sec:validation}).
To measure the global error estimating the masses of this subset of structures, we computed the root mean squared error (rmse), which is defined as
\begin{equation}
\label{eq:rmse}
rmse = \sqrt{\frac{1}{N}\sum_{n=1}^{N}\left(M_{hand}(n) - M_{alg}(n)\right)^2},
\end{equation}
where $M_{hand}(n)$ and $M_{alg}(n)$ are the masses estimated by "hand" and by the algorithm for shell $n$, respectively. Figure \ref{fig:Fig_1}e shows the obtained rmse as a function of $\Delta R$. It is noted that the minimum rmse is obtained for $\Delta R=0.4$ for which the obtained rmse is equal to $3.8 \times 10^{4}$ M$_\odot$.

 We would like to point out that the normalized
error measure, given by
$$rmse_{norm} = \sqrt{\frac{1}{N}\sum_{n=1}^{N}\left(\frac{M_{hand}(n) - M_{alg}(n)}{M_{hand}(n)}\right)^2}$$

was also estimated, but we noted that the impact on the results is mild. In this case we obtain an optimal $ \Delta R=0.3 $ instead of $\Delta R=0.4$. We decided, however, to consider $\Delta R=0.4$ because it is the one that best reproduces the mass values estimated by hand in the sense that the effective number of coincidences between the masses obtained by the algorithm and by hand is higher.

\end{itemize}

\section{Validation of the algorithm}
\label{sec:validation}

Since the goal of this paper is to determine  the kinetic energy stored in  all the Galactic supershell candidates that have  four or three filled quadrants in the recently published catalogue of \citet{sua14}, we need first to be confident that the  masses obtained by the algorithm are reliable.
To test the values yielded by the algorithm, we have measured individually the masses of 95 GSs belonging to the catalogue, 61 of them with four filled quadrants (Group $A$ from hereon) and the remaining 34  with three filled quadrants (Group $B$ from hereon).
It is important to mention that all 95  GSs were randomly selected.  

\begin{figure*}
\centering
\includegraphics[width=16cm]{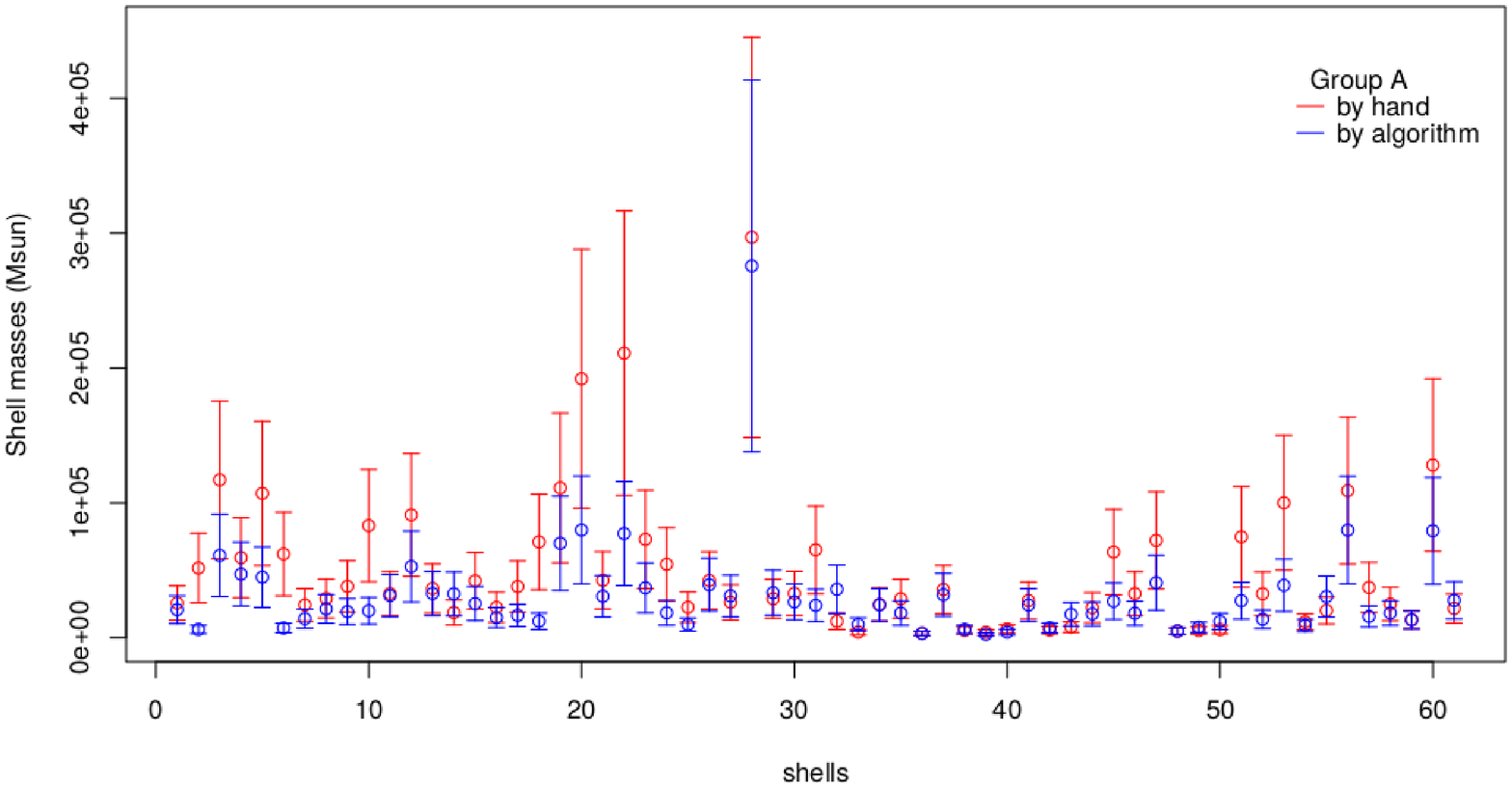}
\includegraphics[width=16cm]{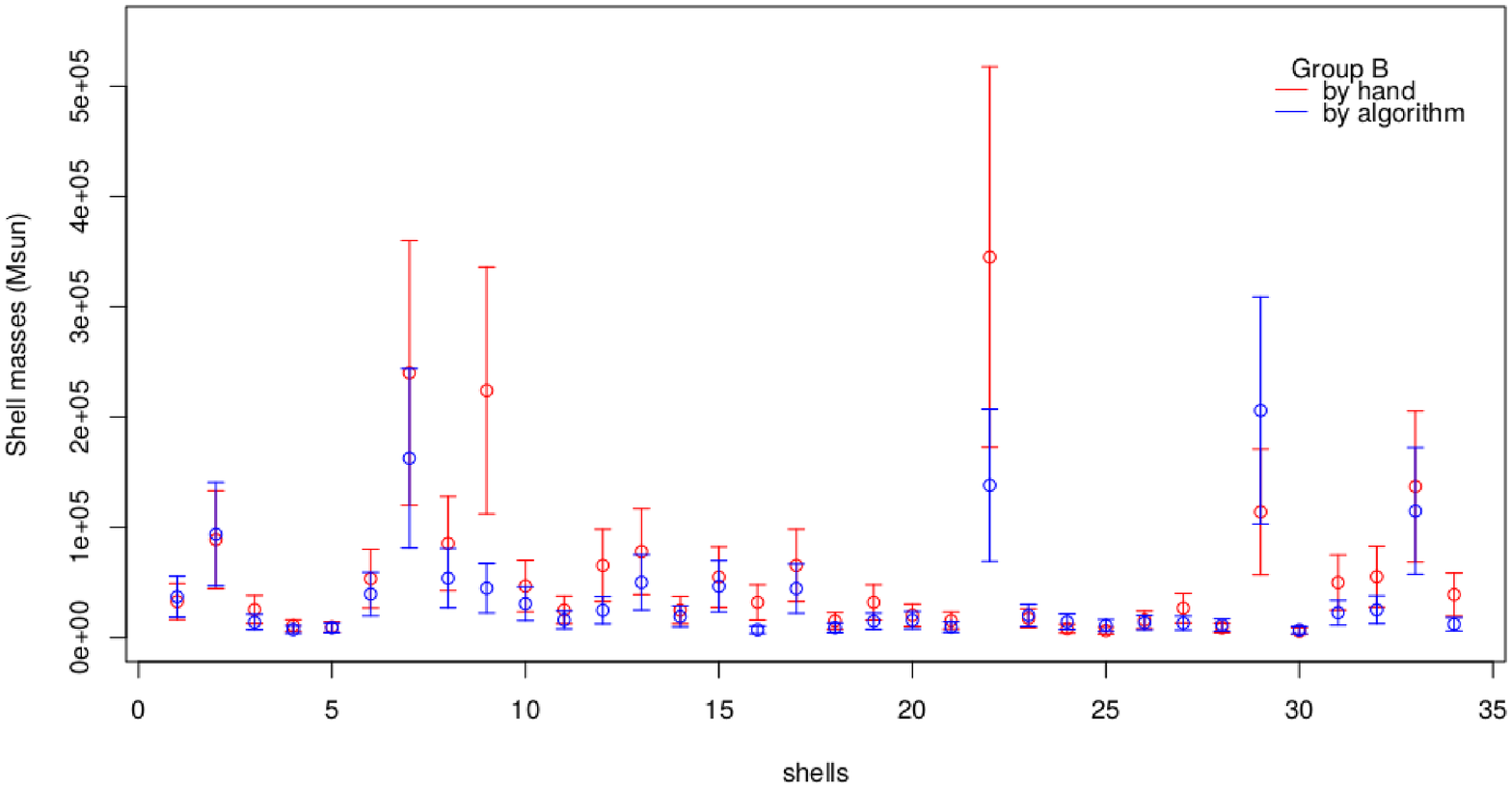}
\caption{Comparison between GS masses obtained by hand (red) and by the algorithm (blue) for  Group $A$  (top panel) and  Group $B$  (bottom panel).
Vertical lines are shown considering a 50 \% error for every mass estimate.} 
\label{3-4C-sc}
\end{figure*}

Following the procedures described in Sects. \ref{hand} and \ref{algorithm}  and using Eq. \ref{eq:masa}, we estimated the 95  GS masses by hand ($M_{\mathrm{\hi\,}}^{\mathrm{shell}}$(Hand)) and by the algorithm ($M_{\mathrm{\hi\,}}^{\mathrm{shell}}$(Alg.)) for structures of Group $A$ and $B$. They are listed in Tables \ref{massesA} and  \ref{massesB}. The distance $d$, and the velocity interval, $\Delta {\rm v}_{\rm km\,s^{-1}}$, where each GS is visible were taken from the catalogue presented in \citet{sua14}.
Figure \ref{3-4C-sc} shows a comparison between the shell masses, $M_{\mathrm{\hi\,}}^{\mathrm{shell}}$,  obtained by hand and by the algorithm, for Group $A$ and Group $B$.
Assuming an error of 50\% for all the estimations, the values obtained from both procedures agree in 93\% (Group $A$) and 91\% (Group $B$) of the structures.

To analyse  the reasons for which the algorithm does not work correctly for some GSs,
we have inspected in detail those GSs that show mass discrepancies.
We found  that for three  (GS 2, 10, and 18; see Table \ref{massesA}) and two (GS 9 and 16; see Table \ref{massesB}) GSs belonging to Groups $A$ and $B$, respectively,  the problem arises from the fact that the algorithm detects emission in the interior cavity of the shell. This could be attributed to  emission  originated by gas not related to the GS, that is, background or foreground emission. As a consequence, the algorithm "sees" a smaller structure and yields a lower mass estimate.
Another detected problem, found for GS 6 (see Table \ref{massesA}) and GS 34 (see Table \ref{massesB}), was that the algorithm fails in detecting the maxima defining the GSs because the \hi\, emission associated with the GS is weak and does not differ from the background emission.

\begin{table}

\caption{GS masses comparison  for Group \it {A}. } 
\label{massesA}
\centering
\begin{tabular}{l c c c }
\hline \hline
 
   \,\,\, &ID & $M_{\mathrm{\hi\,}}^{\mathrm{shell}}$ (Hand)  &  $M_{\mathrm{\hi\,}}^{\mathrm{shell}}$ (Alg.)  \\
  &  &  $\times 10^4$\, M$_\odot$ & $\times 10^4$\, M$_\odot$\\
\hline

1       &       GS093--06--034  &       2.6     &       2.1     \\
2       &       GS100--06--019  &       5.2     &       0.6     \\
3       &       GS101--02--037  &       12.0    &       6.1     \\
4       &       GS101+29--026   &       5.9     &       4.7     \\
5       &       GS102--08--054  &       11.0    &       4.5     \\
6       &       GS104+03--038   &       6.2     &       0.7     \\
7       &       GS105--12--040  &       2.4     &       1.4     \\
8       &       GS107+02--069   &       2.9     &       2.1     \\
9       &       GS107+13--040   &       3.8     &       1.9     \\
10      &       GS108--03--022  &       8.3     &       2.0     \\
11      &       GS108+03--088   &       3.3     &       3.1     \\
12      &       GS113--01--075  &       9.1     &       5.3     \\
13      &       GS113--14--042  &       3.6     &       3.3     \\
14      &       GS114--03--054  &       1.9     &       3.2     \\
15      &       GS114--05--062  &       4.2     &       2.5     \\
16      &       GS115--05--054  &       2.2     &       1.5     \\
17      &       GS118+01--044   &       3.8     &       1.6     \\
18      &       GS119--04--058  &       7.1     &       1.2     \\
19      &       GS121--05--037  &       11.0    &       7.0     \\
20      &       GS122--02--077  &       19.0    &       8.0     \\
21      &       GS124--09--043  &       4.2     &       3.1     \\
22      &       GS129+05--061   &       21.0    &       7.7     \\
23      &       GS133--07--045  &       7.3     &       3.7     \\
24      &       GS135--09--056  &       5.4     &       1.8     \\
25      &       GS136--09--033  &       2.2     &       1.0     \\
26      &       GS137+03--063   &       4.2     &       3.9     \\
27      &       GS138+02--053   &       2.6     &       3.1     \\
28      &       GS140--03--079  &       30.0    &       28.0    \\
29      &       GS141--10--042  &       2.9     &       3.3     \\
30      &       GS144--03--054  &       3.3     &       2.7     \\
31      &       GS144+08--031   &       6.5     &       2.4     \\
32      &       GS146--11--025  &       1.2     &       3.6     \\
33      &       GS146--11--045  &       0.4     &       1.0     \\
34      &       GS153+02--047   &       2.5     &       2.4     \\
35      &       GS164+00--021   &       2.9     &       1.8     \\
36      &       GS195+28+014    &       0.3     &       0.3     \\
37      &       GS198--01+035   &       3.6     &       3.2     \\
38      &       GS199--13+025   &       0.5     &       0.6     \\
39      &       GS201--23+025   &       0.4     &       0.2     \\
40      &       GS202+10+014    &       0.6     &       0.4     \\
41      &       GS221--03+045   &       2.8     &       2.4     \\
42      &       GS222+13+026    &       0.6     &       0.7     \\
43      &       GS227+05+051    &       0.8     &       1.7     \\
44      &       GS229+03+073    &       2.2     &       1.8     \\
45      &       GS230--06+040   &       6.4     &       2.7     \\
46      &       GS232+02+081    &       3.3     &       1.8     \\
47      &       GS239--02+068   &       7.2     &       4.1     \\
48      &       GS240+00+035    &       0.5     &       0.5     \\
49      &       GS240+05+033    &       0.5     &       0.8     \\
50      &       GS246+07+048    &       0.6     &       1.2     \\
51      &       GS247+00+086    &       7.5     &       2.7     \\
52      &       GS253--12+053   &       3.2     &       1.4     \\
53      &       GS253+07+062    &       10.0    &       3.9     \\
54      &       GS256--16+055   &       1.2     &       0.9     \\
55      &       GS257+00+067    &       2.0     &       3.0     \\
56      &       GS259--08+090   &       11.0    &       8.0     \\
57      &       GS260--04+081   &       3.7     &       1.6     \\
58      &       GS261--03+055   &       2.5     &       1.8     \\
59      &       GS263+10+020    &       1.3     &       1.3     \\
60      &       GS265--06+082   &       13.0    &       7.9     \\
61      &       GS269+04+044    &       2.2     &       2.8     \\
\hline
\end{tabular}
\end{table}

\begin{table}
\caption{GS  masses comparison  for Group \it {B}.} 
\label{massesB}
\centering  
\begin{tabular}{l c c c }
\hline \hline
\,\,\, &ID & $M_{\mathrm{\hi\,}}^{\mathrm{shell}}$ (Hand)  &  $M_{\mathrm{\hi\,}}^{\mathrm{shell}}$ (Alg.)  \\
  &  &  $\times 10^4$\, M$_\odot$ & $\times 10^4$\, M$_\odot$ \\
\hline

1       &       GS089--21--025* &       3.3     &       3.7     \\
2       &       GS093--14--021* &       8.9     &       9.4     \\
3       &       GS093+11--034*  &       2.5     &       1.4     \\
4       &       GS098--25--018* &       1.1     &       0.7     \\
5       &       GS098+24--032   &       1.0     &       0.9     \\
6       &       GS100+09--040*  &       5.3     &       3.9     \\
7       &       GS101--13--056* &       24.0    &       16.0    \\
8       &       GS103+07--018*  &       8.5     &       5.4     \\
9       &       GS105--03--061  &       22.0    &       4.5     \\
10      &       GS108+00--075*  &       4.7     &       3.1     \\
11      &       GS109+06--032   &       2.5     &       1.6     \\
12      &       GS109+16--033*  &       6.5     &       2.5     \\
13      &       GS110--04--067  &       7.8     &       5.0     \\
14      &       GS116--06--042* &       2.5     &       1.9     \\
15      &       GS117+08--076*  &       5.5     &       4.7     \\
16      &       GS120+08--028   &       3.2     &       0.7     \\
17      &       GS120+16--067*  &       6.5     &       4.4     \\
18      &       GS130+00--068*  &       1.5     &       0.9     \\
19      &       GS139+06--054*  &       3.2     &       1.5     \\
20      &       GS142--01--057  &       2.0     &       1.6     \\
21      &       GS153--10--026* &       1.6     &       1.0     \\
22      &       GS161+03--036   &       34.0    &       14.0    \\
23      &       GS202+05+031    &       1.7     &       2.0     \\
24      &       GS218--05+037*  &       0.8     &       1.4     \\
25      &       GS224--18+036   &       0.6     &       1.1     \\
26      &       GS240--13+064*  &       1.6     &       1.3     \\
27      &       GS246--05+086   &       2.7     &       1.3     \\
28      &       GS247+06+055*   &       0.9     &       1.1     \\
29      &       GS252--04+074*  &       11.0    &       21.0    \\
30      &       GS257--25+030*  &       0.6     &       0.7     \\
31      &       GS257+09+037    &       5.0     &       2.3     \\
32      &       GS262--09+048*  &       5.5     &       2.5     \\
33      &       GS263--08+068*  &       14.0    &       11.0    \\
34      &       GS264--04+044   &       3.9     &       1.2     \\
\hline
\\
\end{tabular}
\end{table}
  
  \section{Results}
  
Given that, as shown in Sect. \ref{sec:validation}, the algorithm  works correctly in more than 90\% of the structures used to test it, we can now use it to estimate the masses  of the 490 GS candidates belonging to the \cite{sua14}  catalogue. Among them, 308  are completely surrounded by walls of \hi\, emission (Group $A$) and in the remaining 182 the central \hi\, minimum is surrounded by ridges of \hi\, emission in at least 270$^\circ$ of its angular extent (Group $B$).

\begin{figure*}
\centering
\includegraphics[width=16cm]{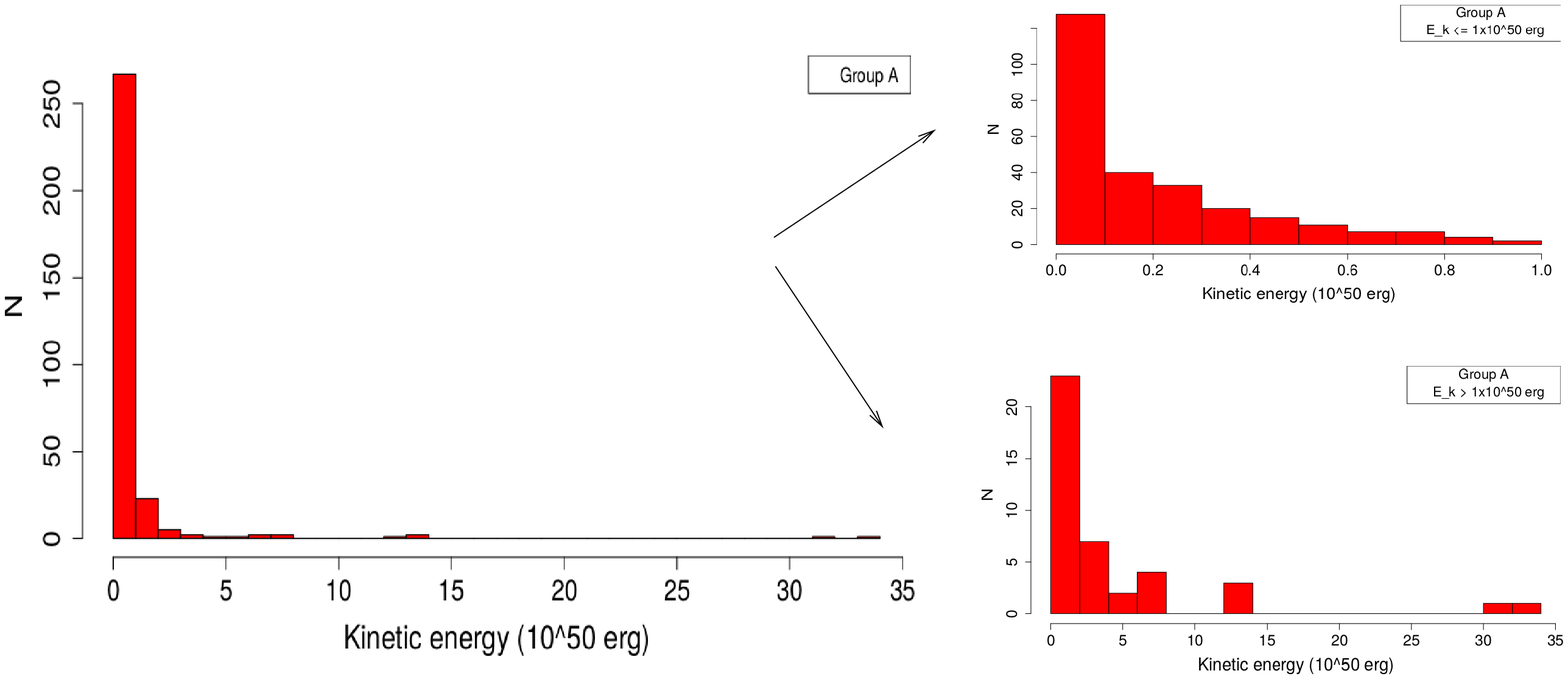}
\includegraphics[width=16cm]{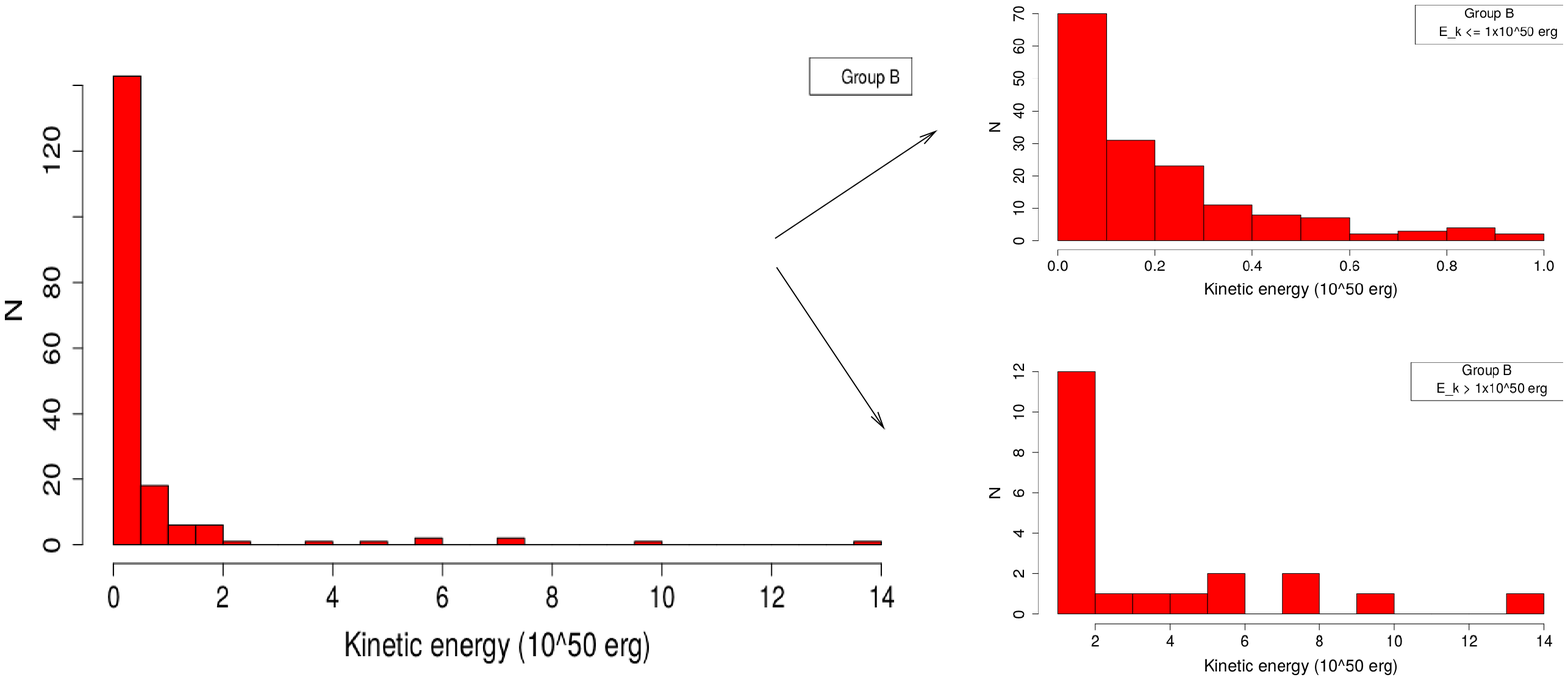}
\caption{Kinetic energy distributions for Group $A$ and  $B$ in top and bottom panels, respectively. Right panels display the energy distributions in more detail by showing structures with  two energy ranges: lower and upper $1\times10^{50}$ erg.} 
\label{hist-Energy}
\end{figure*}

Adopting  solar abundances, the total gaseous mass of each GS is given by

\begin{equation}
M_{\mathrm{t}}^{\mathrm{shell}}=1.34 \, M_{\mathrm{\hi\,}}^{\mathrm{shell}}
;\end{equation}

\noindent then, GS kinetic energies can be derived from

\begin{equation}
E_k = 0.5 \, M_{\mathrm{t}}^{\mathrm{shell}} \, v_{\rm exp}^2
,\end{equation}

\noindent where $v_{\rm exp}$ is the GS expansion velocity
and  is taken to be equal to half the velocity interval where the structure is observed.
The results are shown in Tables 3 and 4 for Groups $A$ and $B$, respectively. The estimated errors are about 50\% for the GS masses and 64\% for the kinetic energies. 
 It is important to mention that although the expansion velocity of an \hi\, expanding structure is usually estimated as half the velocity interval where the structure is detected, this value is an upper limit of the actual expansion velocity, since the differential rotation of the Galaxy should be considered, especially for large structures, as was done by \citet{mcc06} and \citet{ehl18}.
In this work, since we are dealing with 490 structures and many of them are located  at high Galactic latitudes or large distances, where the available rotation models of the Galaxy are probably not adequate, we  estimate the energies without considering the velocity gradient of the Galaxy but with the caution of knowing that   the actual energy values may be lower. 

To check to what extent the effect of the differential rotation affects the energies obtained, we have estimated the energy values using the rotation model of \citet{bra93}. As a result we found that if we call $f$ the quotient between the new expansion speed value and the previous one $f = v^{\rm new}_{\rm exp} /v_{\rm exp}$, considering the 490 structures, $f$ has an average value of 0.75. As for the new values of the energies, we found that for  85\% of them they agree, within errors, with the previous estimated values.

\begin{figure*}
\centering
\includegraphics[width=8cm]{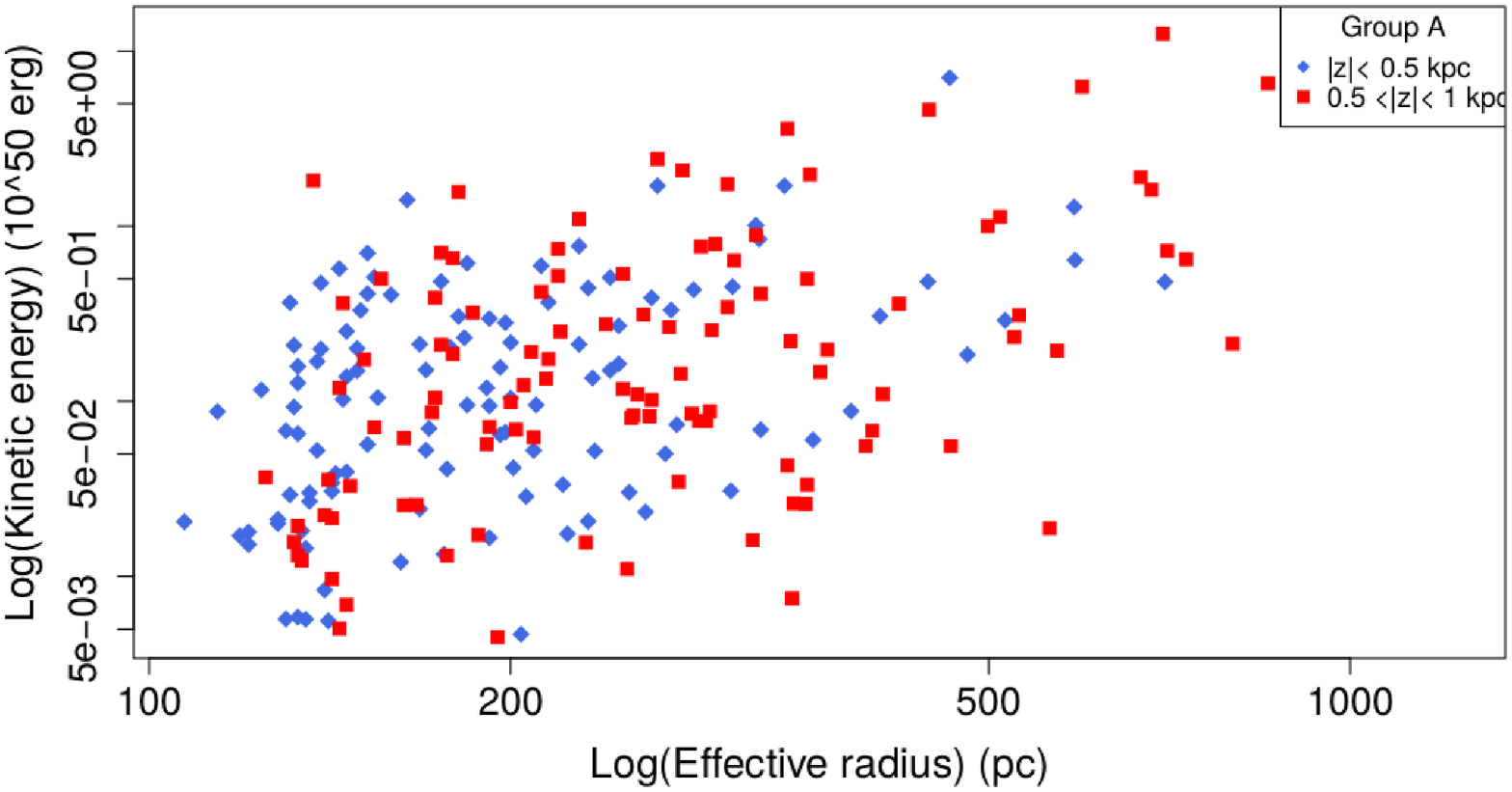}
\includegraphics[width=8cm]{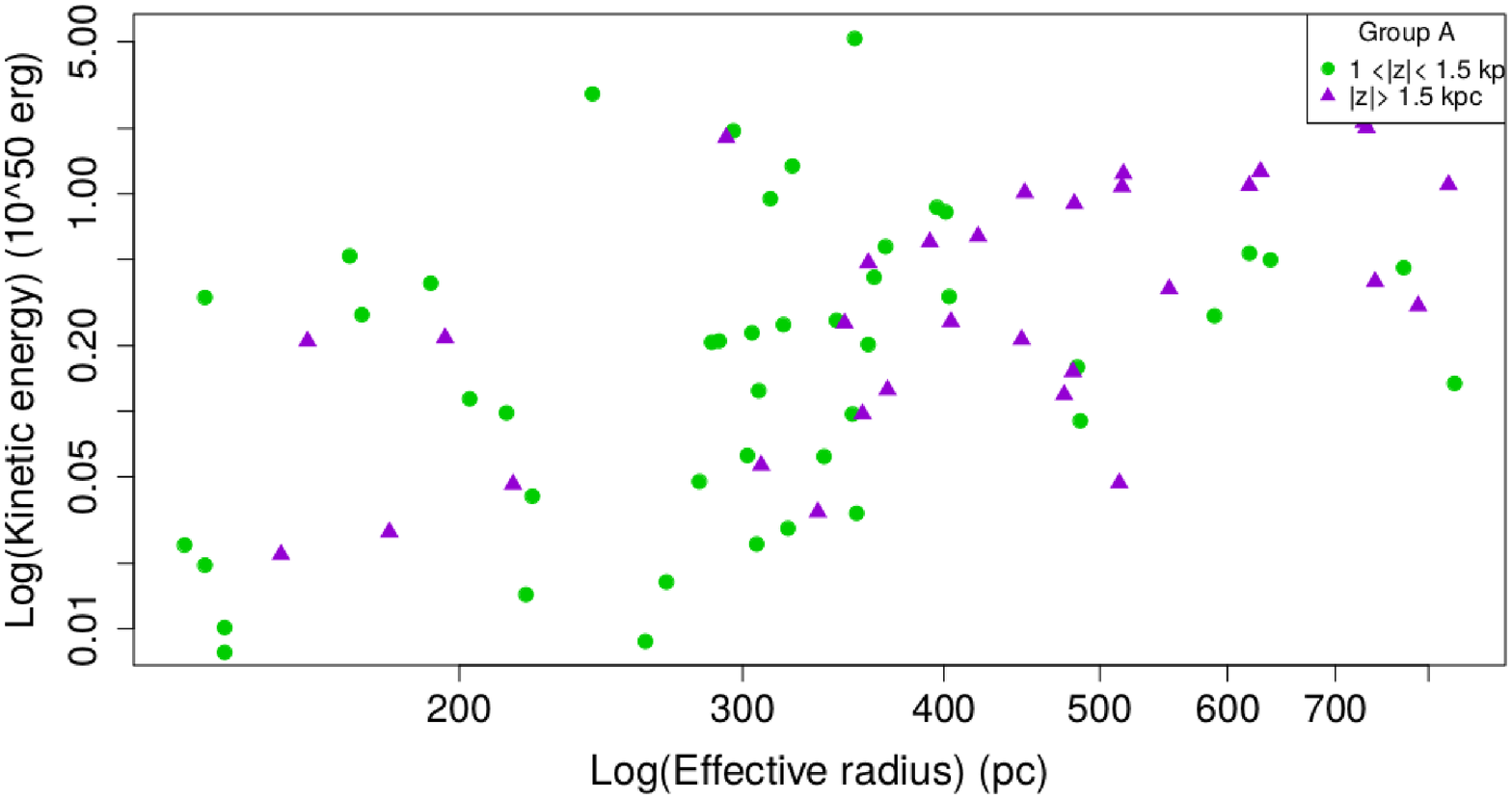}
\includegraphics[width=8cm]{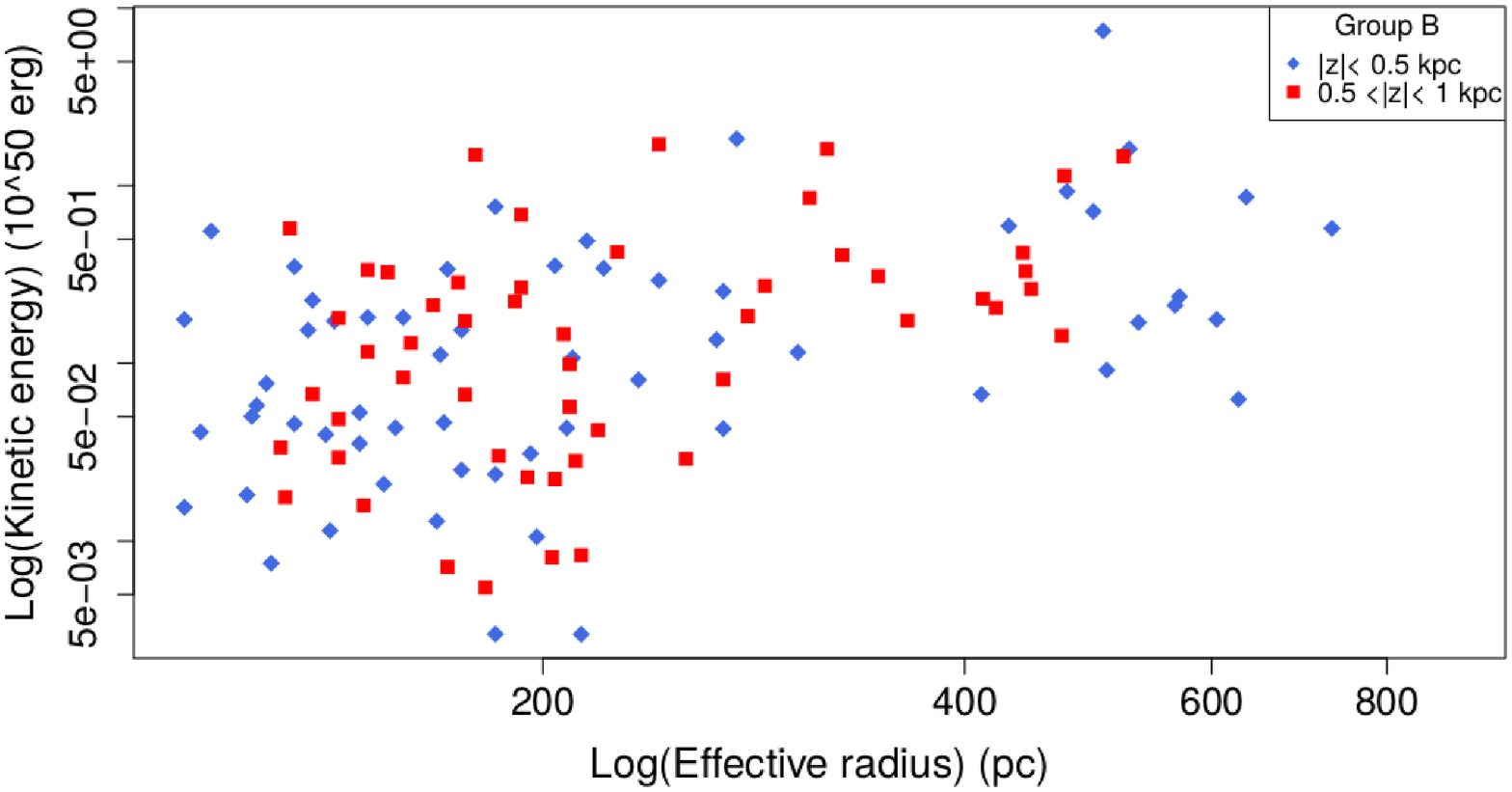}
\includegraphics[width=8cm]{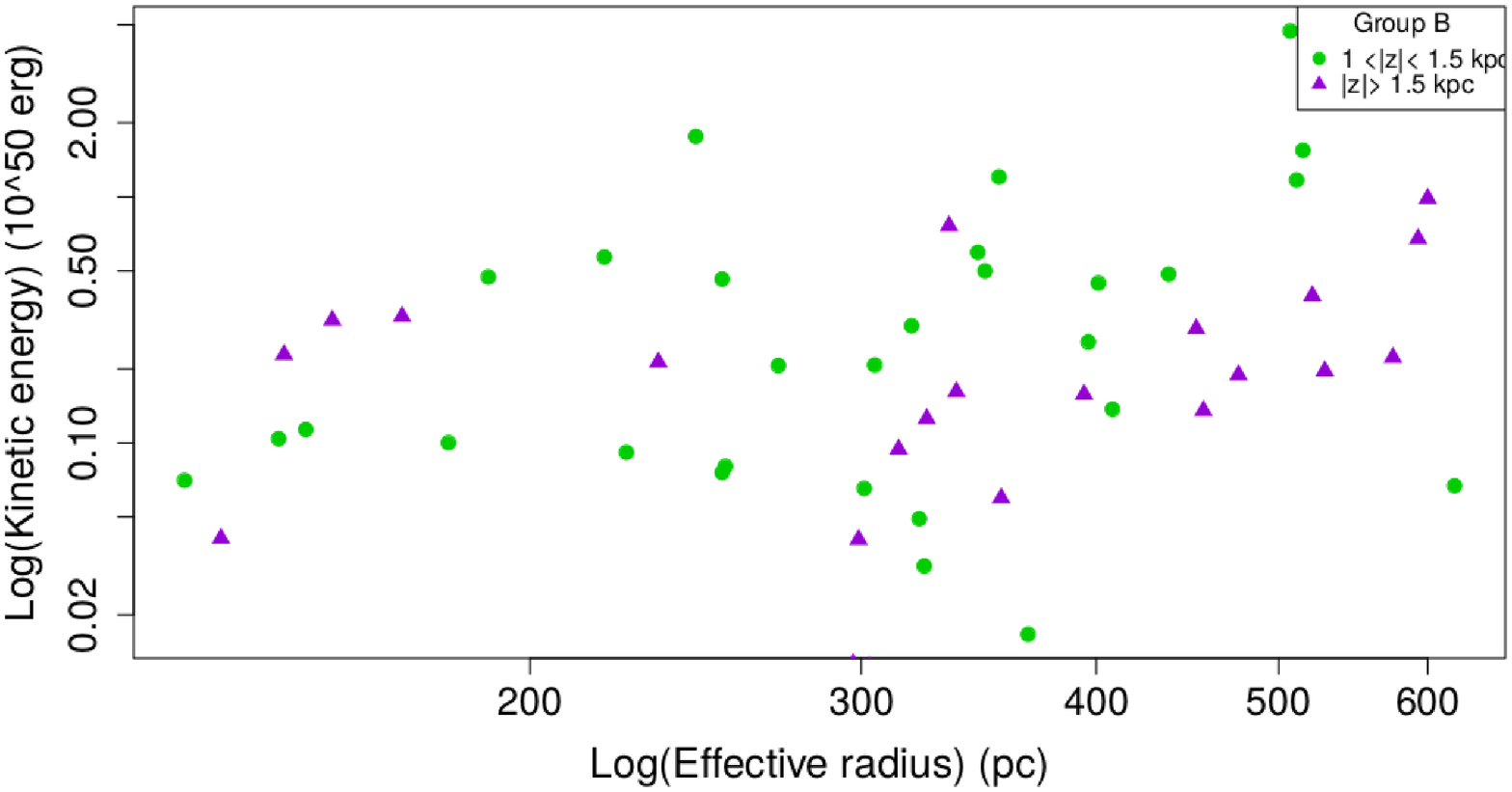}
\caption{Kinetic energy versus effective radius of the GSs for Group $A$ (top    panels) and $B$ (bottom panels). Different  coloured symbols represent different Galactic plane heights. Blue diamonds: |z|$ < 0.5$ kpc, red squares: 0.5 < |z| < 1 kpc, green circles: 1 < |z| < 1.5 kpc, and violet triangles: |z| >1.5 kpc .}
\label{Reff-Energy-Lat}
\end{figure*}

As can be seen in the histograms presented in Fig. \ref{hist-Energy}, we find that the  kinetic energies are between $2.5 \times 10^{47}$ erg and  $3.4 \times 10^{51}$ erg for Group $A$, and between $1 \times 10^{47}$ erg and  $1.4 \times 10^{51}$ erg for Group $B$. Although we found some structures that have very high energies, the mean values are $8 \times 10^{49}$ erg and  $6 \times 10^{49}$ erg for Groups $A$ and $B$, respectively.
Moreover, we found that for Group $A$($B$), 77\% (79\%) of the GSs have energies lower than $0.5 \times  10^{50}$ erg, and 94\% (95\%) lower than $2 \times 10^{50}$ erg.
Figure \ref{Reff-Energy-Lat}  shows a plot of the kinetic energies stored in the GSs versus their effective radius (Reff), considering different Galactic plane
heights (z). 
We find a slight tendency in the relationship between the size and energy, in the sense that the larger structures have higher energies, only for structures closer to the Galactic plane (left panels of Fig. \ref{Reff-Energy-Lat}).
Most of the structures (72\% and 64\% for Groups $A$ and $B$, respectively)  are located at |z| < 1 kpc (see both left panels).
The mean kinetic energy and effective radius for structures belonging to  Group $A$ and located at |z| $\le$ 1 kpc are $0.5 \times10^{50}$ erg and 256 pc, respectively. For GSs located at |z| $>$ 1 kpc the mean values of energy and effective radius are  $1.5 \times10^{50}$ erg and 441 pc.
Regarding  Group $B$, the mean values are $0.5 \times 10^{50}$ erg and 259 pc for |z| $\le$ 1 kpc, and $0.8 \times10^{50}$ erg and 433 pc for |z| $>$ 1 kpc.

\begin{figure*}
\centering
\includegraphics[width=8cm]{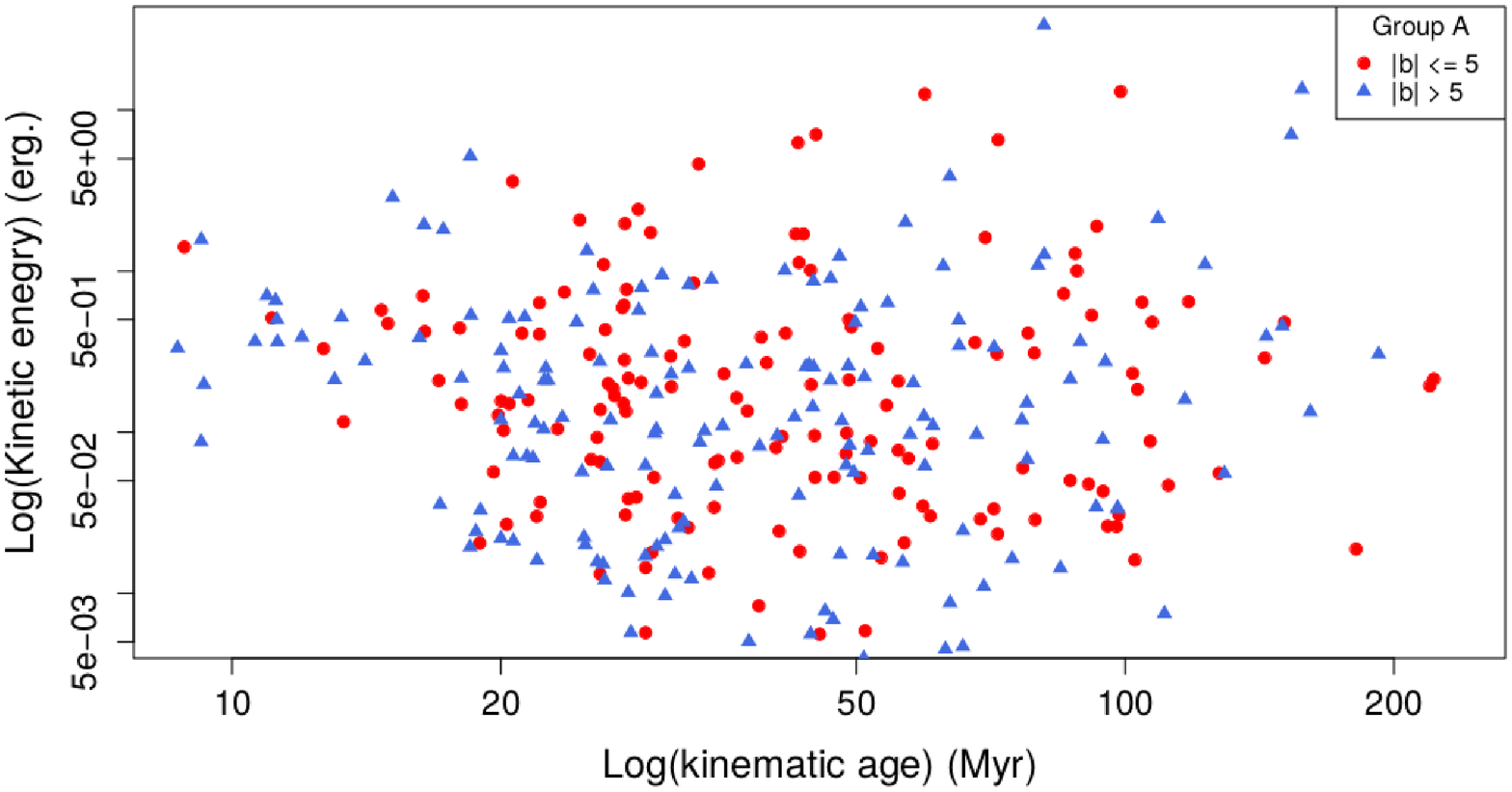}
\includegraphics[width=8cm]{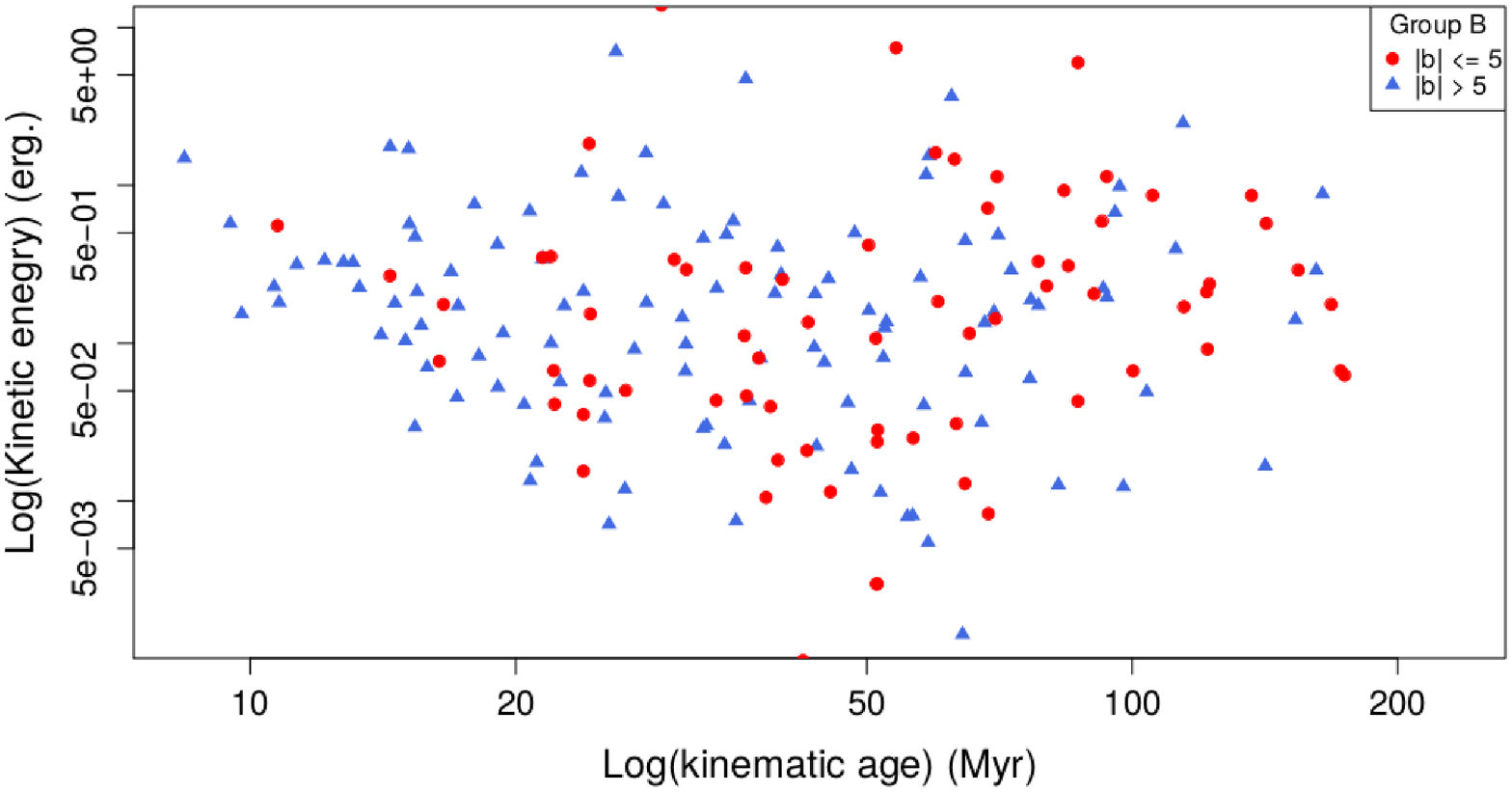}
\caption{Kinetic energies versus kinematic ages for GSs at different Galactic latitudes. Red points:  GSs located at |b| $\le$ 5$^\circ$, blue triangles: GSs located at |b| > 5$^\circ$.}
\label{Energy-age}
\end{figure*}

Figure \ref{Energy-age} shows a plot of the  kinetic energy of the GSs versus their kinematic ages, for Groups $A$ (left panel) and $B$ (right panel), considering different Galactic latitude intervals (indicated by different colours).
The kinematic ages  were taken from the GS candidates catalogue \citep{sua14}, where  they were  obtained by  using   $t\,(\rm{Myr}) = Reff\,\rm{(pc)}/v_{exp}\,$\rm{(km\,s}$^{-1}$).  Bearing in mind that the assumed expansion velocities are upper limits, the ages obtained are lower limit values (they are, on average, a factor of 1.4 larger if the $v^{\rm new}_{\rm exp}$ is considered).
No clear dependence between kinetic energy and kinematic age is detected. Both groups  have similar mean kinematic ages, 49 $\pm$ 12 Myr (Group $A$) and 53 $\pm$ 13 Myr (Group$B$). 
 It is important to note, however, that these estimated ages decrease  if we assume that the GSs were created by the action of stellar winds, in which case their dynamical ages are given by $t\,(\rm{Myr}) = 0.6\, Reff\,\rm{(pc)}/v_{exp}\,$\rm{(km\,s}$^{-1}$) \citep{wea77}.

Another important parameter that can be derived from the estimated GS masses is the ambient density ($n_0$) of the ISM local to each structure. 
By uniformly distributing  the excess  mass in the shell  over the cavity where the mass is missing, the ambient density into which a GS is evolving can be derived as

\begin{equation}
n_0 = 10 \, \frac{M_{\mathrm{t}}^{\mathrm{shell}}}{\rm vol} \hspace{5cm}  [\rm {cm}^{-3}]
,\end{equation}

\noindent where $M_{\mathrm{t}}^{\mathrm{shell}}$ is in solar masses and "vol" is the volume of the cavity in pc$^3$, and was calculated assuming a spherical geometry. 
These estimates were done only for the 95 GSs whose masses were computed by the algorithm and by hand, because the volume of individual cavities is not easy to define using the algorithm. 
We obtain $n_0$ values going from 0.03 to 2.5 cm$^{-3}$.

\begin{figure}
\centering
\includegraphics[width=8cm]{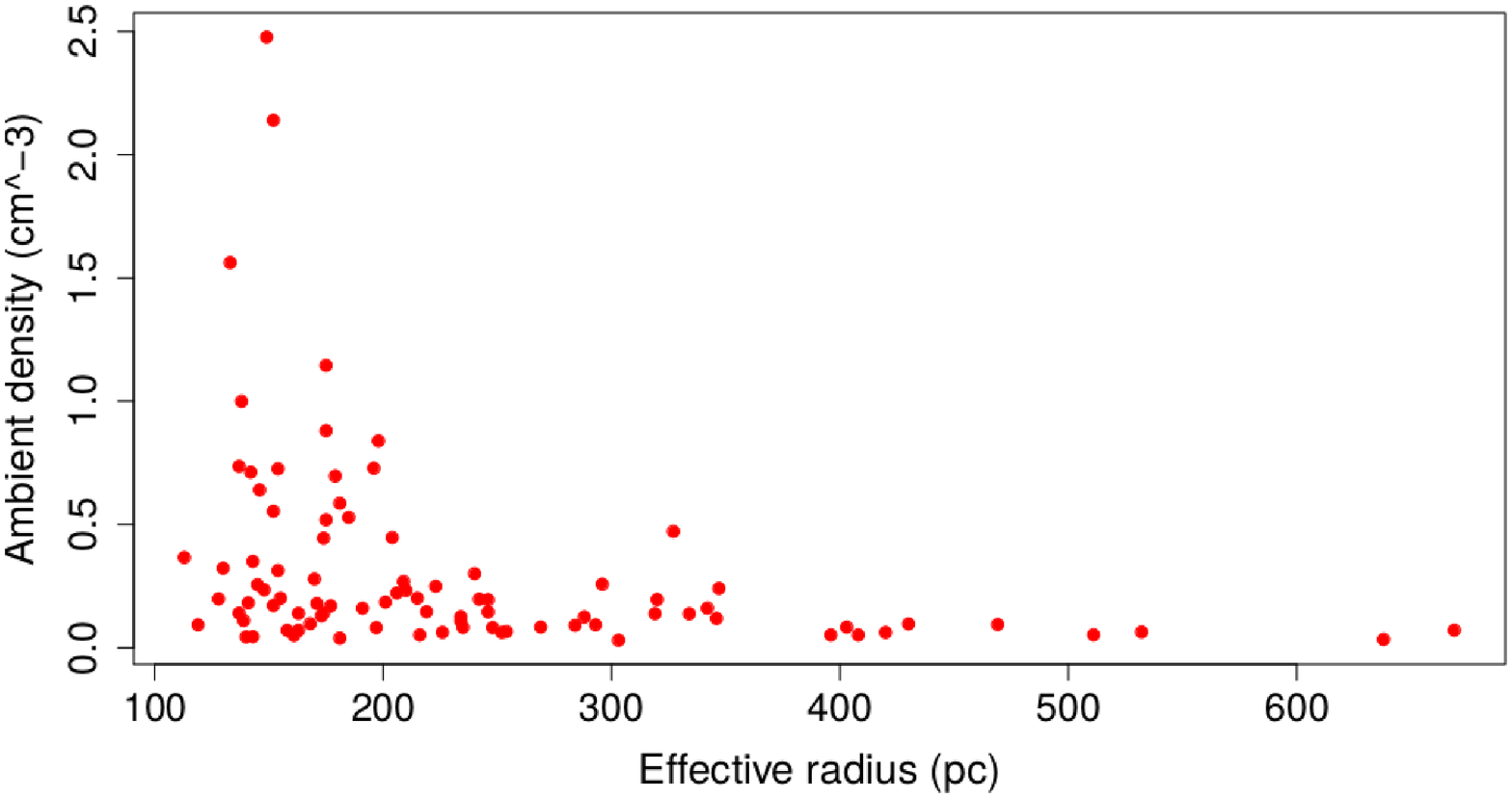}
\caption{Effective radius versus ambient density for the subsample of 95 GSs. }
\label{noR}
\end{figure}

Figure \ref{noR} shows the relation between $n_0$ and the effective radius of the GSs. It can be seen that the largest structures seem to be located in regions with lower ambient densities. We found that for GSs with an effective radius higher than 250 pc, the averaged ambient density is 0.25 cm$^{-3}$, while for the rest it goes up to 0.72 cm$^{-3}$.

\begin{figure}
\centering
\includegraphics[width=8cm]{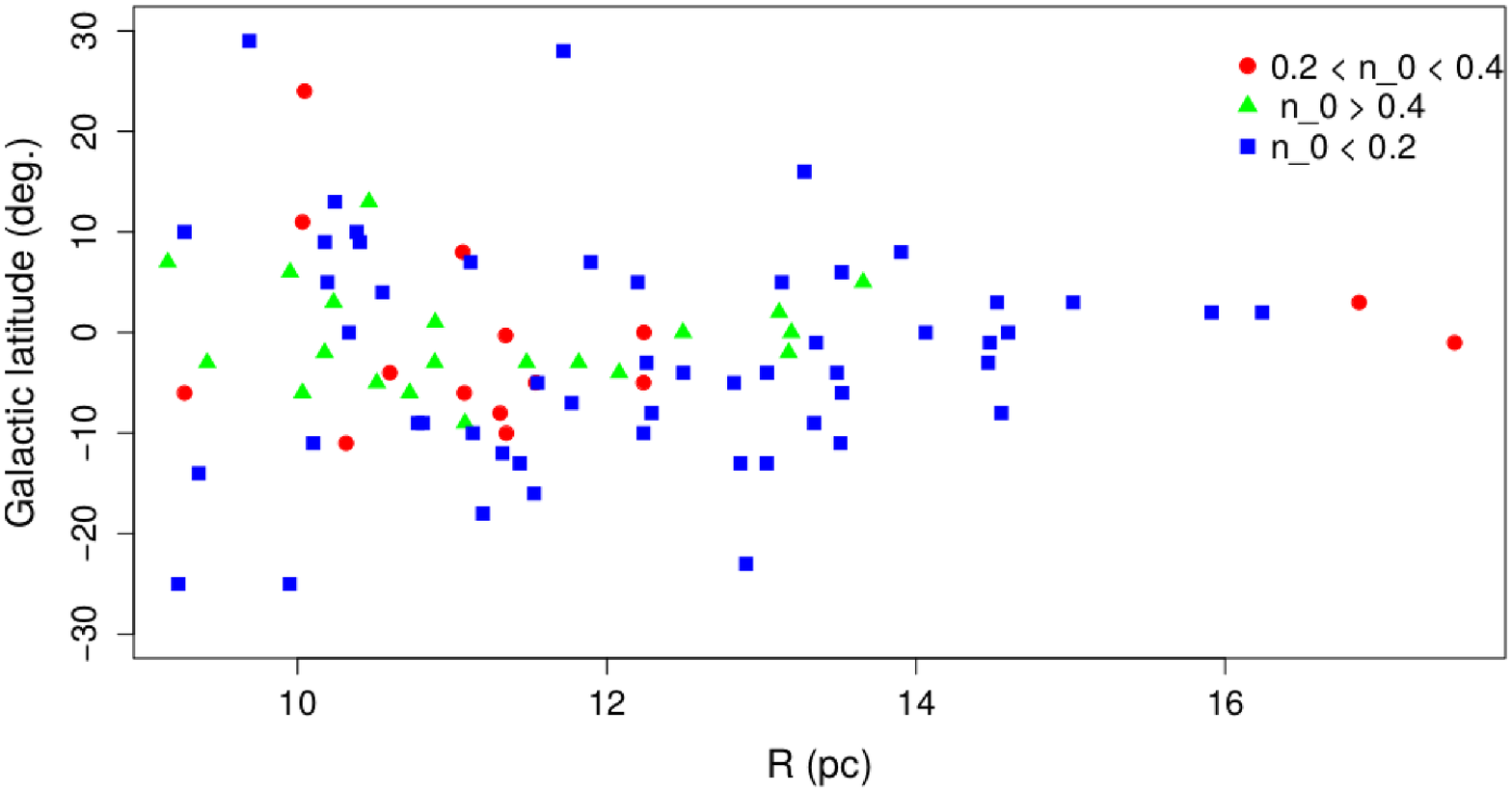}
\caption{Ambient density values distribution. Red dots represent ambient densities between 0.2 and 0.4 cm$^{-3}$, green triangles and blue squares represent densities above 0.4 cm$^{-3}$ and lower than 0.2 cm$^{-3}$, respectively.}
\label{density}
\end{figure}

On the other hand, to analyse the dependence of the estimated ambient densities with respect to the GS's location in the Galaxy, in Fig. \ref{density} we plot them in a graph of Galactic latitude versus  Galactocentric distance (R), considering three  ambient density intervals. 
Ambient density values lower than 0.2 cm$^{-3}$, between 0.2 and 0.4 cm$^{-3}$, and  greater  than 0.4 cm$^{-3}$  are represented with  blue, red, and green dots, respectively.

From Fig. \ref{density} it is clear  that, as expected, ambient densities greater than 0.4 cm$^{-3}$ are located mostly between $\pm$ 10 degrees Galactic latitude and not beyond 14 kpc from the Galactic centre. Lower densities are located in a wider range of Galactic latitudes and  Galactocentric distances.

\section{Discussion}

As mentioned in Sect. \ref{intro}, the action of massive stars is the most probable mechanism for the formation of shell structures. 
Taking into account the theoretical model of \citet{wea77}, the efficiency of conversion of mechanical stellar wind energy $E_w$ into kinetic energy $E_k$ is up to $20\%$ for the energy conserving model. As was mentioned above, most of the structures has energies lower than $2 \times 10^{50}$ erg, thus the wind energy needed to create a GS with this energy is $E_w \sim 1 \times 10^{51}$ erg. 
These values can be perfectly reached by stellar OB associations. For example an OB association with two O6.5\,V, two O7\,V, and three O8\,V stars that contribute,  during their main sequence (MS) phase,   $E_w \sim 1.1 \times 10^{51}$ erg, could explain the origin of the GS. In the Galactic  OB  Associations  in  the Northern Milky Way Galaxy  \citep{gar92}, most of the OB associations fulfil this requirement. Open clusters, such as NGC\,6193, are also capable of injecting the required wind energy.

Nevertheless, apart from the required energies, to analyse the origin of the GSs we have to take into account their ages.
 The fact that  we found that most of the GSs are older than 
typical OB associations  suggests that, for the formation of each structure, more than one generation of massive stars should be involved and that contributions from one or more supernova (SN) explosions are to be expected. 
Since, for example, the MS  lifetime of an O7 type star is 6.4 Myr \citep{sch92}, the earlier type stars, in an OB association,  are expected to 
be evolved or  have already exploded as a SN.
Given that,  assuming a stellar origin for the GS, only 19  are younger than 6.5 Myr, we do not expect to detect earlier type stars than an O7\,V in the interior of  most of them. On the other hand, taking into account that the SN rate in a typical OB association is about one per 10$^5$ - 3$\times 10^5$ yr \citep{mcc87}, over the lifetime of the GSs it is reasonable to assume that the most massive stars may have exploded.

For example, for one of the catalogued GSs, GS\,100$-$02$-$041, \cite{sua12}  suggested  that  the  wind energy provided during the  main-sequence phase of the now evolved massive stars
belonging to the OB association Cep\,OB1, and located inside GS\,100$-$02$-$41, could explain the origin of the \hi\, supershell, whose kinetic energy was estimated as $1.8\times 10^{50}$ erg. However, given the advanced age of the structure (5.5 Myr), they do not discard the possibility that a supernova explosion could have taken place.

 Assuming that the genesis of the majority of the GSs was deeply rooted to the
stellar winds and supernova explosions of massive stars, the location of the GS
in the outer part of the Milky Way would suggest that in the past massive
stars were located there.

\begin{figure}
\centering
\includegraphics[width=8cm]{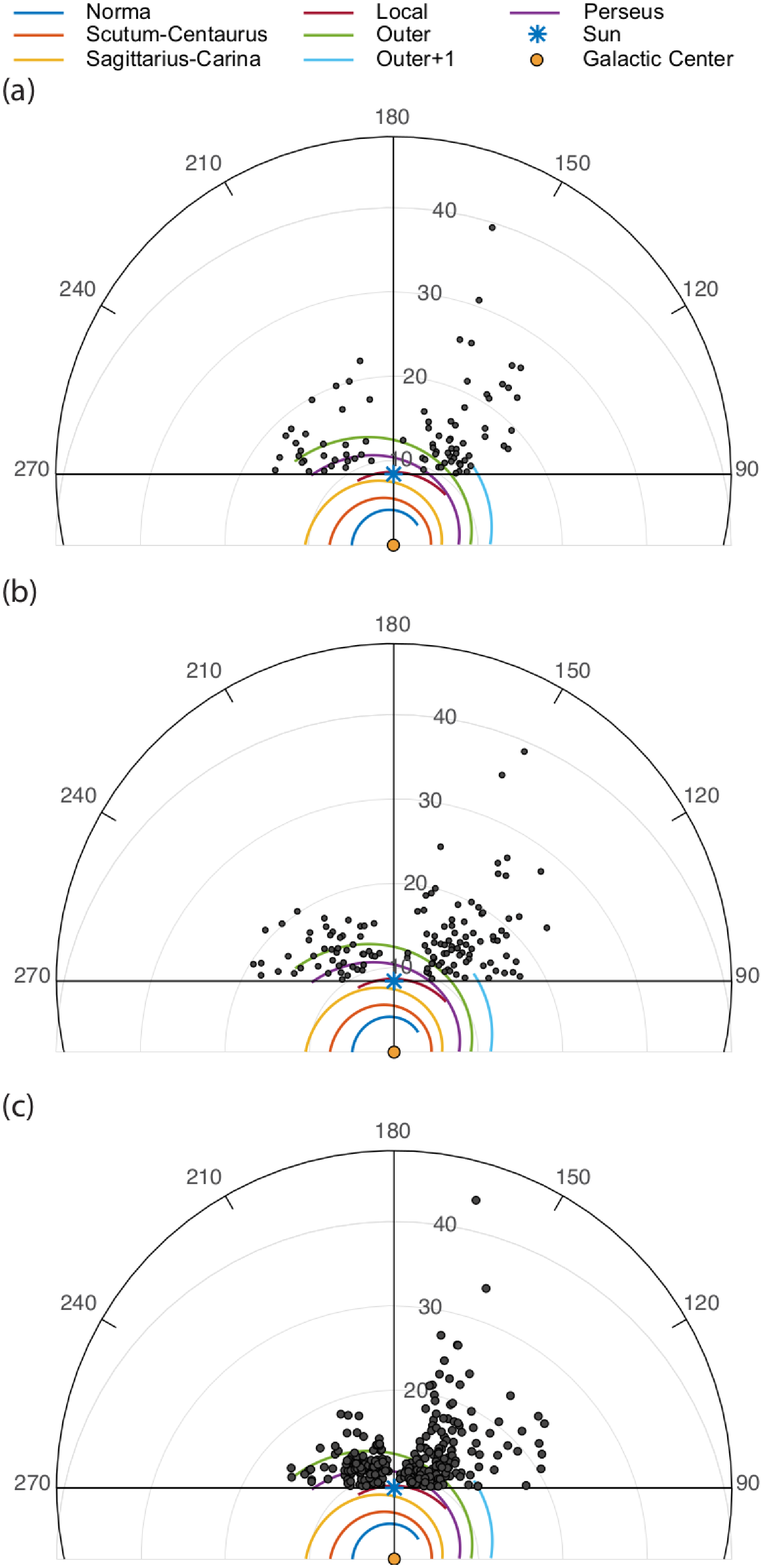}
\caption{Best-fitted models of polynomial-logarithmic (PL) spirals \citep{hou14}. Black points represent the GSs.  Panel  (a) shows the GSs located at $|b|<= 2^\circ$. Panel (b): GSs located at $2^\circ < |b| < 5^\circ$. Panel (c): GSs located at $|b| >= 5^\circ$. The Sun is represented by a blue asterisk and the Galactic center by a yellow dot. The concentric circles around the Galactic center are separated by 10 kpc and indicate the distance to the Galactic center.}
\label{Galactic-distribution}
\end{figure}

\begin{figure}
\centering
\includegraphics[width=8cm]{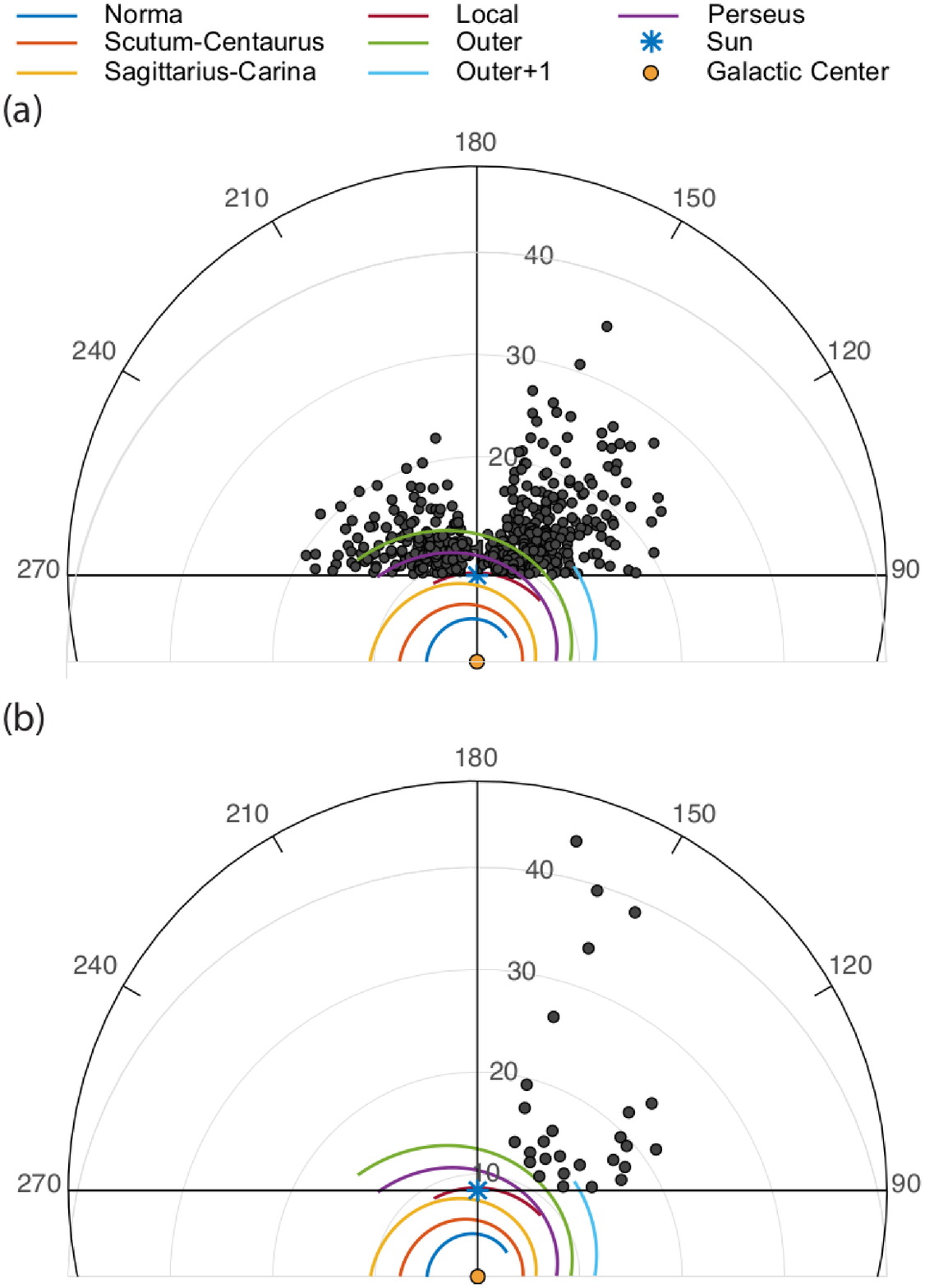}
\caption{Same galactic model as in Fig. \ref{Galactic-distribution}. Black points represent the GSs. In  (a) and (b) panels are plotted the GSs with $E_k$ $\le$ 2$\times 10^{50}$ erg and $E_k$ > 2$\times 10^{50}$ erg, respectively.}
\label{Galactic-energy-distribution}
\end{figure}

The GSs' spatial distribution in the outer part of the Milky Way is shown
in Fig. \ref{Galactic-distribution}, whilst the location of GSs, depending on their kinetic energy, is shown in Fig. \ref{Galactic-energy-distribution}. To plot the spiral arms of the Galaxy, we used the best-fitted models of polynomial-logarithmic (PL) spirals derived by \cite{hou14} using three kinds of spiral tracers (H\,II regions, giant molecular clouds, and masers). At first glance it is striking in both figures that a sizable fraction of the GSs present in the second Galactic quadrant are located at much larger Galactocentric distances than those located in the third quadrant. This finding was also pointed out by \cite{ehl13} in their analysis,  where they also detected this asymmetry between the second and third Galactic quadrants in the distribution of shells at large Galactocentric  distances. The origin of this observational result could be attributed to the gas distribution beyond the furthest spiral arm. Recent studies 
indicate that M31 has an extended gaseous halo \citep{leh15}, suggesting that \hi\, structures located at large distances could be formed in the circumgalactic medium of the Andromeda galaxy (M31) and in the intra-group gas in the Local Group filament, as was analysed in a recent paper by \cite{ric17}. 

On the other hand, as can be seen in  Fig. \ref{Galactic-distribution}, most of the GSs located at $|b| \le  2^\circ$ (upper panel)  in the third Galactic quadrant seem to be projected onto the Perseus Arm  and the Outer Arm.   In this Galactic quadrant only a few GSs appear beyond the Outer Arm. At the extreme end of the Outer+1 Arm  we can see  a cumulus of several GSs. 
In the middle panel of Fig. \ref{Galactic-distribution},  GSs located at $2^\circ< |b| < 5^\circ$  are plotted. As in the   upper panel, there are  several GSs delineating the Perseus and  Outer Arms  but, in this case, there are more GSs located beyond the Outer Arm. Beyond the Outer+1 Arm, GSs appear  to be more dispersed than in the upper panel. Finally, in the   bottom panel of  Fig. \ref{Galactic-distribution}  GSs located at $|b| \ge 5^\circ$ are plotted. In the second Galactic quadrant there is an enhancement of GSs beyond the Outer Arm. Between $165^\circ < l < 195^\circ$ a lack of GSs is in evidence.  This is because this region of the sky  was not considered for \cite{sua14} to look for supershells, since the rotation curve used to derive kinematic distances is not trustworthy there.

We can conclude that GSs seem to be formed, at least for $|b|\le 2^\circ$, in the Galactic spiral arms, which is consistent with a stellar origin.
However, as shown in Fig. \ref{Galactic-distribution}, there are  some GSs  located at higher Galactic latitudes, showing that  if they also were formed by massive stars, these stars are or were present in this region of the Galaxy.
As mentioned in Sect. \ref{intro}, another possible origin for these GSs is  collision with a HVC. As shown in Fig. 2 of \cite{ric17}, there are a lot of HVC located at high latitudes, so it could be possible that some HVC played a role in the origin of some GSs.

The fact that  we found GSs located  at large Galactocentric  distances, $R > 15 $ kpc, may seem strange  or inconsistent with a stellar origin since, until now, it was believed that massive stars were mostly located in the spiral arms.
Nevertheless, a recent study  of young star-forming regions associated with  molecular clouds revealed that many of them are found at $R \ge  13.5 $ kpc, in the outer Galaxy \citep{izu17}.  They detected 711 new candidate star-forming regions in 240 molecular clouds. 
On the other hand, \cite{and15} detected H\,II regions located at Galactocentric  distances greater than 19 kpc at $l \sim 150^\circ$. Galactic H\,II regions are the formation sites of massive OB stars.
These results  strongly show that there is star formation activity in regions beyond the Outer and Outer+1 Arms. 
Concerning large \hi\, structures, \cite{cic11} detected a GS, GSH\,91.5+2$-$114, in  the outer part of the Galaxy, located at a distance of about 15 kpc from the Sun. Based on an analysis of the energetics and of the main physical parameters of the large shell,  they conclude that GSH\,91.5+2$-$114 is likely to be the result of the combined action of the stellar winds and supernova explosions of many stars.

Finally, Fig. \ref{Galactic-energy-distribution} shows the location of the GSs with kinetic energies lower than 2$\times 10^{50}$ ergs (Fig. \ref{Galactic-energy-distribution}a) and  greater than  2$\times 10^{50}$ erg (Fig. \ref{Galactic-energy-distribution}b). In Fig. \ref{Galactic-energy-distribution}a, 
the GSs  appear more
uniformly distributed in the second and third Galactic quadrants than in Fig. \ref{Galactic-energy-distribution}b, with a clear concentration towards the spiral arms.  On the contrary, Fig. \ref{Galactic-energy-distribution}b shows that the most energetic structures  are mostly  located beyond the Outer and Outer+1 Arms.

\section{Conclusions}

In this paper, we have estimated the amount of \hi\, mass of each supershell candidate belonging to the \cite{sua14} catalogue using an automatic algorithm, which was tested by a comparison of the results it gives with the ones obtain by hand, for 95 structures. 
 For the analysis we divided the GSs into two groups: 
those with four filled quadrants belong to Group $A$  and those with three filled quadrants to Group $B$.

The analysis of the obtained  energies allows us to draw the following conclusions:

-- 95 \% of the supershells have kinetic energies lower than $2 \times 10^{50}$ erg, which, for a stellar origin, implies that a wind energy greater than $1 \times 10^{51}$ erg is required. 

-- There is no clear correlation between the energy stored in a shell and its  kinematic age. 
With respect to the linear size, we find that the size increases accordingly as the energy increases only in those structures located near the Galactic plane.

-- Although there is no clear difference in the energy values found for Groups $A$ and $B$, we found out that for |z| > 1 kpc,  the mean kinetic energy for GSs belonging to Group $A$ is larger than the one found for Group $B$. This may indicate that kinetic energy is being lost in the open structures.  

-- According to the interval of energies obtained, a stellar origin is possible for most of the GSs. However, more than one star is needed and, in most of the cases, given that their ages are not extremely young, more than one stellar generation, including several SN explosions.

-- An origin due to a collision with a HVC is also possible, especially for those GSs located at very high latitudes.

-- The GSs found at very high distances from the Galactic centre may be formed by the diffuse gas 
 that connects the stellar body of our Galaxy with  the Local Group  environment.

In summary, we conclude that most of the large \hi\, structures found in the outer part of the Galaxy could have been created by the action of several generations of massive stars and SN explosions. 
If this is actually the case, the GSs may be used to look for massive stars not yet detected  and/or to better understand the massive stellar formation and distribution history.

\begin{acknowledgements}
 This work was partially financed by  grants PIP\,0226 and  PIP\,0604 of the Consejo Nacional de Investigaciones Cient\'{i}ficas  y  T\'ecnicas  (CONICET)  of  Argentina.  We would like to thank the anonymous referee for
constructive and useful comments that helped us to considerably improve the
quality of the original paper.             
\end{acknowledgements}

\bibliographystyle{aa} 
\bibliography{bibliografia}
 
 \IfFileExists{\jobname.bbl}{}
{\typeout{}
\typeout{****************************************************}
\typeout{****************************************************}
\typeout{** Please run "bibtex \jobname' ' to optain}
\typeout{** the bibliography and then re-run LaTeX}
\typeout{** twice to fix the references!}
\typeout{****************************************************}
\typeout{****************************************************}
\typeout{}
}

\end{document}